\documentclass[]{JHEP3}
\usepackage{epsfig}
\def\be{\begin{equation}}
\def\ee{\end{equation}}
\def\bea{\begin{array}}
\def\eea{\end{array}}
\def\beqa{\begin{eqnarray}}
\def\eeqa{\end{eqnarray}}
\def\beqas{\begin{eqnarray*}}
\def\eeqas{\end{eqnarray*}}

\def\bp{\begin{picture}}
\def\ep{\end{picture}}
\def\bc{\begin{center}}
\def\ec{\end{center}}
\def\bfig{\begin{figure}}
\def\efig{\end{figure}}

\def\bit{\begin{itemize}}
\def\eit{\end{itemize}}
\def\nn{\nonumber}
\def\f{\frac}

\def\[{\left[}
\def\]{\right]}
\def\({\left(}
\def\){\right)}

\def\..{\left.}
\def\.{\right.}
\def\tl{\tilde}
\def\ra{\rightarrow}
\def\la{\leftarrow}

\def\tm{\times}

\def\da{\dagger}

\def\la{\lambda}

\def\al{\alpha}

\def\ka{\kappa}

\def\ep{\epsilon}

\def\pa{\partial}
\def\pr{\prime}

\title{NMSSM with generalized deflected mirage mediation}
\author{Xiao Kang Du$^{1}$, Guo-Li Liu$^1$, Fei Wang$^{1}$, Wenyu Wang$^{2}$,  Jin Min Yang$^{3}$, Yang Zhang$^4$\\
$^1$ School of Physics, Zhengzhou University, 450000, ZhengZhou
P.R.China\\
$^2$ School of Physics, Beijing University of Technology, Beijing, P.R.China\\
$^3$ CAS Key Laboratory of Theoretical Physics, Institute of Theoretical Physics, Chinese Academy of Sciences,
     Beijing 100190, China\\
$^4$ ARC Centre of Excellence for Particle Physics at the Tera-scale, School
of Physics and Astronomy, Monash University, Melbourne, Victoria 3800,
Australia \\
}

\abstract{
We propose to generate a realistic soft SUSY breaking spectrum for Next-to-Minimal Supersymmetric Standard Model (NMSSM) with a generalized deflected mirage mediation scenario,
 in which additional Yukawa and gauge mediation contributions are included to deflect the renormalization group equation(RGE) trajectory. Based on the Wilsonian effective action obtained by integrating out the messengers, the NMSSM soft SUSY breaking spectrum can be given analytically at the messenger scale. We find that additional contributions to $m_S^2$ can possibly ameliorate the stringent constraints from the electroweak symmetry breaking (EWSB) and 125 GeV Higgs mass. Constraints from dark matter and fine-tuning are also discussed.
 The Barbieri-Giudice fine-tuning measure and electroweak fine-tuning measure
  in our scenario can be as low as ${\cal O}(1)$, which possibly indicates that
  our scenario is natural.

}

\begin{document}
\maketitle

\newpage
\section{Introduction}
Low energy supersymmetry (SUSY), which can solve elegantly the gauge hierarchy problem of the Standard Model (SM)
and provide a viable dark matter (DM) candidate, has been regarded by many physicists as one of the most appealing candidates for the TeV-scale new physics.
However, the reported data of the Large Hadron Collider (LHC) agree quite well with the SM predictions
and no significant deviations have been observed in the electroweak precision measurements
and the flavor physics. Besides, the lack of SUSY signals at the LHC \cite{CMSSM1,CMSSM2} and
the difficulty to accommodate the discovered 125 GeV Higgs \cite{ATLAS:higgs,CMS:higgs} seem to indicate
that the low energy SUSY spectrum should display an intricated structure.
As the low energy soft SUSY breaking spectrum is determined by the SUSY breaking mechanism,
it is interesting to survey the phenomenology related to the SUSY breaking and mediation mechanism
from a top-down approach.

 SUSY breaking mechanism from flux compactification of type IIB string theory can lead to interesting consequences.
In the generalized Kahler-modulus dominated scenarios, the dilaton and complex structure moduli fields
can be stabilized by the background NS and RR 3-form fluxes. Such superheavy modes will be integrated out and eliminated from the low energy effective theory. The remaining Kahler moduli fields can be stabilized by non-perturbative effects, such as the gaugino condensation.
To generate SUSY breaking in the observable sector and tune the cosmological constant to
a tiny positive value, Kachru et al propose to add an anti-D3 brane at the tip of the Klebanov-Strassler
throat to explicitly break SUSY and lift the AdS universe to obtain a dS one \cite{KKLT}.
Consequently, the F-component of the light Kahler moduli fields could mediate the SUSY breaking effects
to the visible sector and result in a mixed modulus-anomaly mediation SUSY breaking
scenario \cite{mirage:hep-th:0411066,mirage:hep-th:0503216}. It is interesting to note that the involved
modulus mediated SUSY breaking contributions can be comparable to that of the anomaly mediation \cite{AMSB}.
With certain assumptions on the Yukawa couplings and the modular weights, the change of the SUSY breaking contributions (from the inputs) by the renormalization group equation(RGE) evolution and the anomaly mediation contributions could cancel each other at a $'mirage'$
scale, leading to an apparent pure modulus mediation scenario at such a mirage scale\cite{mirage:hep-ph:0504037}.
Such a mixed modulus-anomaly mediation SUSY breaking mechanism is dubbed as $'mirage~mediation'$.

Anomaly mediation contribution is a crucial ingredient of such a mixed modulus-anomaly mediation mechanism. It is well known that purely anomaly mediation mechanism is bothered by the tachyonic slepton problem \cite{tachyonslepton}. One of its non-trivial extensions with a messenger sector, namely the deflected anomaly mediation SUSY breaking (AMSB) scenario, can elegantly solve such a tachyonic slepton problem through the deflection of the RGE trajectory \cite{dAMSB,okada,fei} by additional gauge\cite{GMSB} or Yukawa mediation contributions. Such a messenger sector can also be present in the mirage mediation scenarios and play an important role
in generating a preferable low energy SUSY spectrum. In fact, the $'mirage'$ unification of gaugino masses at TeV scale can be possible in deflected mirage mediation scenarios, even with the simplest minimal KKLT set-up\cite{deflectmirage}. In the mixed modulus-anomaly mediation mechanism, a straightforward extension of $\mu$ sector to predict a small $B\mu$ term needs some fine tuning \cite{mirage:hep-ph:0504037}. Such a $\mu-B\mu$ problem can be solved naturally in the framework of Next-to-Minimal Supersymmetric Standard Model(NMSSM). The realization of NMSSM in TeV mirage mediation scenarios had already been discussed in the literature \cite{mirageNMSSM1,mirageNMSSM2}.
However, it was found that only a small portion of the parameter space can be consistent with
the electroweak symmetry breaking(EWSB) conditions and at the same time accommodate the 125 GeV Higgs \cite{mirageNMSSM1}.
So it is rather interesting to seek new ways to generate a realistic NMSSM spectrum.

Additional gauge or Yukawa mediation contributions in the mirage mediation \cite{deflectmirage} can deflect the RGE trajectory and change the low energy soft SUSY predictions. Analytical expressions for the soft SUSY breaking parameters at the messenger scale are given in
the Wilsonian effective action approach \cite{1808.08529} by one of the authors.
Based on the general discussions in \cite{1808.08529}, we propose to generate the NMSSM spectrum by a generalized deflected mirage mediation
mechanism \cite{gdmirage} with additional gauge and Yukawa mediation contributions.
We find that the inclusion of the messenger sector and non-trivial interactions can possibly alleviate the stringent constraints from the 125 GeV Higgs and the EWSB conditions.

This paper is organized as follows. We briefly review the mirage mediation mechanism in Section \ref{sec-2}.
We state our model and present the analytical expressions for the soft SUSY parameters in Section \ref{sec-3}.
The numerical results are discussed in Section \ref{sec-4}.
Section \ref{sec-5} contains our conclusions.
\section{\label{sec-2} Brief review of the mirage mediation mechanism}

In Type IIB string theory compactified on a Calabi-Yau orientifold, the presence of background fluxes can fix the dilaton and the complex structure
moduli, leaving only the Kahler moduli in the four-dimensional Wilsonian effective supergravity action (defined at the boundary scale $\Lambda$)
after integrating out the superheavy complex structure moduli and dilaton.
The low energy effective Lagrangian in terms of compensator field and a single Kahler modulus that parameterizes the overall size of the
compact space\cite{mirage:hep-ph:0504037} is given as
\begin{eqnarray}
e^{-1}{\cal L}=\int d^4\theta \[\phi^\da\phi \(-3 e^{-K/3}\)-(\phi^\da\phi)^2\bar{\theta}^2\theta^2 {\cal P}_{lift}\]
+\int d^2\theta \phi^3 W+ \int d^2\theta \f{f_i}{4} W_i^a W_i^a,
\label{action}
\end{eqnarray}
with a holomorphic gauge kinetic term
\beqa
f_i=\f{1}{g_i^2}+i\f{\theta}{8\pi}.
\eeqa

The Kahler potential involves the $'no-scale'$ kinetic form for the Kahler modulus while
the superpotential involves the KKLT setup
\beqa
W=\(\omega_0-A e^{-aT}\)+W_0~,
\label{KKLT:superpotential}
\eeqa
where the first term is generated from the fluxes and the second term from non-perturbative effects,
such as gaugino condensation from the non-abelian gauge sector or D3-brane instantons. $W_0$ denotes the superpotential terms involving the MSSM(NMSSM) sector as well as a possible messenger sector. The modulus $T$, which is not fixed by the background flux, can be stabilized by such a KKLT-type superpotential with
\beqa
a~\Re{\langle T \rangle}\approx\ln \(\f{A}{\omega_0}\)\approx \ln\(\f{M_{Pl}}{m_{3/2}}\)\approx 4\pi^2~,
\eeqa
up to ${\cal O}(\ln[{M_{Pl}}/{m_{3/2}}]^{-1})$. Note that in the KKLT setup, the flux-induced SUSY breaking is dynamically canceled by the non-perturbative SUSY breaking that stabilizes the Kahler modulus $T$, leading to a SUSY-preserving AdS solution.
In order to obtain a vacuum with a positive cosmological constant and break SUSY,
KKLT proposed to add an $\bar{D}3$ brane to provide an uplifting operator given by
\beqa
{\cal P}_{lift}=D (T+T^\da)^{n_P}~,
\eeqa
with a positive constant $D\sim{\cal O}(m_{3/2}^2 M_{pl}^2)$.
The uplifting operator, which represents the low energy consequence of the
sequestered SUSY-breaking brane, is independent of visible matter fields
and $T$ (with $n_P=0$) in the minimal KKLT set-up.
Explicit SUSY breaking via anti-D3 branes can be replaced by typical D-term or F-term 
uplifting mechanisms \cite{F-D-lifting}.

 With the uplifting low energy effective potential, we have the leading order F-terms 
 of compensator field $F_\phi$ and Kahler modulus $F_T$\cite{mirage:hep-ph:0504037}
 \beqa
&& F_\phi \approx  m_{3/2}\approx\f{\omega_0}{M_{Pl}^2(T+T^*)^{3/2}}~, \\
&& M_0\equiv\f{F_T}{T+T^*} \approx \f{2}{a(T+T^*)}m_{3/2}\approx\f{m_{3/2}}{\ln\(\f{M_{Pl}}{m_{3/2}}\)}~.
  \eeqa
The non-zero F-term VEV of the heavy moduli $H$ are given approximately
 by $F_H\sim m_{3/2}^2/m_U\ll m_{3/2}$. Therefore, it gives negligible contributions to SUSY breaking \cite{0804.4283}.

In the mirage mediation, we have $ m_{3/2}\approx (4\pi^2)M_0$ numerically, which means that the loop induced
anomaly mediation contributions are comparable to the modulus mediation contributions.
The importance of the anomaly mediation contributions relative to the modulus mediation contributions
can be parameterized by
\beqa
\label{alpha}
\al^\pr\equiv\f{m_{3/2}}{M_0\ln\(M_{Pl}/m_{3/2}\)}.
\eeqa
So $\al^\pr\ll 1$ corresponds to the limit of pure modulus mediation,
while $\al^\pr\gg 1$ corresponds to the pure anomaly mediation.
Although the minimal KKLT predicts $\al^\pr\approx 1$, other values of $\al^\pr\sim {\cal O}(1)$ can be
obtained by assuming proper uplifting operator and proper forms for the potential of Kahler modulus \cite{mirage:hep-ph:0504037}.
So we leave the value $\al^\pr$ as a free parameter in the following discussions.

\section{\label{sec-3} NMSSM with deflected mirage mediation}
The general form of the low energy effective action eqn.(\ref{action})
can be amended to include new ingredients, for example, the NMSSM sector and a messenger sector.
The Kahler potential will include not only the $'no-scale'$ kinetic form for the Kahler modulus but also additional kinetic terms for messenger fields
\beqa
K&=&-3\ln(T+T^\da)+Z_X(T^\da,T) X^\da X+Z_\Phi(T^\da,T)\Phi^\da\Phi\nn\\
 & & +\sum\limits_{i}Z_{P_i,\bar{P}_i}(T^\da,T)\[ P_i^\da P_i+\bar{P}_i^\da \bar{P}_i\]~,
\eeqa
with $\bar{P}_i,P_i$ the messenger superfields and $\Phi$ the NMSSM superfield.
The Kahler metric for matter fields and messengers as well as holomorhic gauge kinetic functions
are assumed to depend non-trivially on the Kahler modulus $T$
\beqa
Z_X(T^\da,T)&=&\f{1}{(T^\da + T)^{n_X}}~,~~Z_\Phi(T^\da,T)=\f{1}{(T^\da + T)^{n_\Phi}}~,~\nn\\
f_i(T)&=& T^{l_i}~,~~~~~~~~~~~~~Z_{P_i,\bar{P}_i}(T^\da,T)=\f{1}{(T^\da + T)^{n_P}}~.
\eeqa
The values of $n_X,n_\Phi,n_P,l_i$ depend on the location of the fields on the D3/D7 branes.
Besides, universal $l_i=1$ are adopted in our scenario so that the gauge fields reside on the D7 brane and the gauge coupling unification can be preserved.

The $W_0$ term within the superpotential eqn.(\ref{KKLT:superpotential}) is given as
\beqa
W_0=W_{\overline{NMSSM}}+W_M~,
\eeqa
with 
\beqa
W_{\overline{NMSSM}}&=&\la S H_u H_d+\f{1}{3}\ka S^3+W_{MSSM}~,
\eeqa
the ($Z_3$ symmetric) NMSSM sector and 
\beqa
W_{M}&=&\sum\limits_{m}\[\la_X^T X \tl{X}_m X_m+\la_X^D X \tl{Y}_m Y_m\]+\la_P^T S \tl{X}_1 X_{2} +\la_P^D S \tl{Y}_1 Y_{2}+ W(X)~,
\eeqa
the messenger sector.
The $'2m'$ family of messengers can be fitted into ${\bf 5,\bar{5}}$ representation of $SU(5)$ gauge group
\beqa
P_m({\bf 5})=X_m(3,1)_{-1/3}\oplus Y_m(1,2)_{~1/2}~,\nn\\
\tl{P}_m({\bf \bar{5}})=\bar{X}_m(\bar{3},1)_{~1/3}\oplus\bar{Y}_m(1,\bar{2})_{-1/2}~.
\eeqa
The VEV of the pseudo-modulus superfield $X$, which can be determined by $W(X)$ and other SUSY breaking effects,
can set the messenger thresholds through the $X P_m\tl{P}_m$ type couplings.
The deflection parameter $'d'$,
which characterizes the relative size of contributions between the gauge (Yukawa) mediation
and anomaly mediation, can be readily obtained
\beqa
d F_\phi\equiv\f{F_X}{X}-F_\phi~.
\eeqa
Similar to the deflected AMSB, a positive deflection parameter in the deflected mirage mediation,
which can be generated by a carefully chosen superpotential etc\cite{okada,fei},
may be preferable because less messenger species are needed
so that the problem of strong gauge couplings below the GUT scale can be evaded.
The purpose to introduce even number of messenger species is to forbid possible kinetic mixing between $X$ and $S$, otherwise the tadpole term for $S$ would destabilize the weak scale \cite{giudice:0706.3873}.
The discrete $Z_3$ breaking by EWSB may generate domain walls in the early universe which may lead to
an unacceptably large anisotropy of CMB. To avoid such a problem, the $Z_3$ symmetry is assumed to be broken by some higher dimensional operators.

There are two approaches to obtain the low energy SUSY spectrum in the deflected mirage mediation:
 \bit
\item  In the first approach, the deflected mirage mediation soft SUSY spectrum is given
by the expressions in \cite{mirage:hep-ph:0504037} at the boundary scale.
      Such a spectrum will receive additional contributions towards its RGE running to low energy scale,
      especially the threshold corrections related to the appearance of messengers\cite{1001.5261}.

\item In the second approach which we will adopt, the soft SUSY spectrum at low energy scale
is derived directly from the (low energy) effective action.
  The SUGRA description in eq.(\ref{action}) can be seen as the Wilsonian effective action after
integrating out the superheavy complex structure moduli and dilaton field.
After the pseudo-modulus $X$ acquires a VEV and determines the messenger threshold, the messengers
can be integrated out to obtain a low energy effective action below the messenger threshold scale.
So we anticipate that the Kahler metric $Z_\Phi$ and gauge kinetic $f_i$ should depend non-trivially
on the messenger threshold $M_{mess}^2/\phi^\da\phi$ and $M_{mess}/\phi$, respectively.
The resulting soft SUSY spectrum below the messenger threshold can be derived from the wavefunction
renormalization approach \cite{GMSB:wavefunction,shih,chacko,fei2}. The analytical expressions for the soft SUSY breaking parameters in the most general form of deflected mirage mediation scenarios are given in \cite{1808.08529} by one of the authors.
\eit
\subsection{The modular weight choices for NMSSM superfields}

We need to specify the modular weights $n_i\equiv 1-m_i$ for NMSSM superfields before we can carry out numerical analysis.
In the NMSSM, successful EWSB and the solution to the $\mu$-problem in general require a large VEV for the
singlet $S$. So it is preferable to introduce a negative $m_S^2$ and/or large trilinear terms  $A_\la,A_\ka$ for the singlet superpotential interactions.
As a negative $m_S^2$ prefers vanishing modulus contributions, we set the modular weight $m_{S}=0$.
 The choice of modulus weight for $H_u,H_d$ can be understood from the EWSB conditions in NMSSM. From
  \beqa
 \f{M_Z^2}{2}&=&\f{m^2_{H_d}-m_{H_u}^2\tan\beta^2}{\tan^2\beta-1}-\mu^2~,~~
  \eeqa
we can see that $m_{H_u}^2$ should be light to avoid a too large fine-tuning.
On the other hand, $m_{H_u}^2\ll m_{H_d}^2$ for $\tan\beta\gg1$. So we can set $m(H_d)=1$ or $1/2$ and $m_{H_u}=0$.

The electroweak(EW) naturalness prefers relatively light stops. In the MSSM, light stops below 1 TeV are disfavored
because it is difficult to accommodate the observed 125 GeV Higgs.
However, large loop corrections involving stops are not necessarily required to interpret the 125 GeV
Higgs in NMSSM. So light stops, which is preferable from the criterion of a low EW fine-tuning, are still allowed in NMSSM.
Squarks of the first two generations should be heavy to avoid various SUSY CP and flavor constraints.
We note that even pure AMSB contributions can guarantee the heaviness of colored SUSY particles.
Besides, it is preferable to ameliorate the $g_\mu-2$ discrepancy in the
framework of SUSY with light sleptons and electroweakinoes. With the modular weight $l_i=1$ for gauge fields, a positive deflection parameter  $'d'$ can possibly guarantee the lightness of the electroweakinos.

 The notorious tachyonic slepton problem in AMSB can be solved in our scenario.
 Positive slepton masses can be obtained by introducing a proper deflection parameter $'d'$ or
 by adding extra modulus mediation contributions. So we chose the following 
$m_i$ in our scenario:
 \bit
\item Modular weights for sleptons are given by $m_{(L_L)^{1,2,3}}=m_{(E_L^c)^{1,2,3}}=1/2$.

\item  Modular weights for other matter and messenger fields are given by
  \beqa
 && m_{H_u}=m_{S}=m_{Q_L^3}=m_{t_L^c}=m_{b_L^c}=0,\nn\\
 && m_{a}=\f{1}{2},~~(a={Q}^{1,2}_L,(U_L^c)^{1,2},(D_L^c)^{1,2})~,\nn\\
 && m_{H_d}=m_{X}=m_{\tl{X}}=m_{Y}=m_{\tl{Y}}=1.
 \label{modular:weight}
  \eeqa
\item  Double messenger species with $m=1$ are adopted in our subsequent numerical study.
  \eit

Note that the messenger modular weights also play a role and contribute to $m_{{S}}^2$.
The modular weights $n_i= 0$ correspond to matter fields on D7 branes while $n_i=1$ on D3 branes.
Modular weights $n_i=1/2$ corresponds to fields on the intersections of the D3-D7 branes.

\subsection{Analytical expressions for soft SUSY breaking parameters}
Now we use the second Wilsonian approach in our analysis. From the analytical expressions for the generalized mirage mediation\cite{1808.08529}, the soft SUSY breaking parameters in NMSSM at the messenger scale $M_{mess}$ after integrating out the messenger fields can be given explicitly.
The gaugino masses are given as
\beqa
M_i=l_i M_0\f{g_i^2(\mu)}{g_i^2(GUT)}+ F_\phi\f{\al_i}{4\pi}\(b_i-d\Delta b_i\)~,
\label{gaugino}
\eeqa
with $l_i=1$ and
\beqa
~(b_1~,b_2~,~b_3)&=&(\f{33}{5},~1,-3)~,\\
\Delta(b_1~,b_2~,~b_3)&=&(~2,~2,~2).
\eeqa
Within the expression, the relative size between the anomaly and modulus mediation contribution
at the messenger scale is determined by the free parameter
\beqa
\al =\f{F_\phi}{(16\pi^2) M_0}.
\label{al:def}
\eeqa
We define the modular weight $q_{y_{ijk}}$ as
\beqa
q_{y_t}&=&m_{Q_{L,3}}+m_{H_u}+m_{t_L^c}~,\nn\\
q_{y_b}&=&m_{Q_{L,3}}+m_{H_d}+m_{b_L^c}~,\nn\\
q_{y_\tau}&=&m_{L_{L,3}}+m_{H_d}+m_{\tau_L^c}~,\nn\\
q_{\la}&=& m_{S}+m_{H_d}+m_{H_u}~,\nn\\
q_{\ka}&=& 3m_{S}~.
\eeqa
The trilinear soft terms at the messenger scale are given by
\small
\beqa
A_t&=&q_{y_t} M_0-\f{M_0}{8\pi^2}\[6y_t^2\f{q_{y_t}}{2}+y_b^2\f{q_{y_b}}{2}+\la^2\f{q_{\la}}{2}-(\f{16}{3}g_3^2+3g_2^2+\f{13}{15}g_1^2)\]\ln\(\f{M_{GUT}}{M_{mess}}\)
\nn\\
&+&\f{F_\phi}{16\pi^2}\[6y_t^2+y_b^2+\la^2-(\f{16}{3}g_3^2+3g_2^2+\f{13}{15}g_1^2)\]~,\\
A_b&=&q_{y_b}M_0
-\f{M_0}{8\pi^2}\[y_t^2\f{q_{y_t}}{2}+6y_b^2\f{q_{y_b}}{2}+\la^2\f{q_{\la}}{2}-(\f{16}{3}g_3^2+3g_2^2+\f{7}{15}g_1^2)\]\ln\(\f{M_{GUT}}{M_{mess}}\)\nn
\\&+&\f{F_\phi}{16\pi^2}\[ y_t^2+6y_b^2+\la^2-(\f{16}{3}g_3^2+3g_2^2+\f{7}{15}g_1^2)\]~,\\
A_\tau&=&q_{y_\tau} M_0-\f{M_0}{8\pi^2}\[3y_b^2\f{q_{y_b}}{2}+\la^2\f{q_{\la}}{2}-(3g_2^2+\f{9}{5}g_1^2)\]\ln\(\f{M_{GUT}}{M_{mess}}\)
\nn\\&+&\f{F_\phi}{16\pi^2}\[3y_b^2+\la^2 -(3g_2^2+\f{9}{5}g_1^2)\]~,\\
A_\la &=&q_{\la} M_0-\f{M_0}{8\pi^2}\[3y_t^2\f{q_{y_t}}{2}+3y_b^2\f{q_{y_b}}{2}+4\la^2\f{q_{\la}}{2}+2\ka^2\f{q_{\ka}}{2}-(3g_2^2+\f{3}{5}g_1^2)\]\ln\(\f{M_{GUT}}{M_{mess}}\)
\nn\\&+&\f{F_\phi}{16\pi^2}\[4\la^2+2\ka^2+3y_t^2+3y_b^2 -(3g_2^2+\f{3}{5}g_1^2)\]+\Delta A_\la~,\\
A_\ka &=& q_{\ka} M_0-\f{M_0}{8\pi^2}\[6\la^2\f{q_{\la}}{2}+6\ka^2\f{q_{\ka}}{2}\]\ln\(\f{M_{GUT}}{M_{mess}}\)
\nn\\&+&\f{F_\phi}{16\pi^2}\[6\la^2+6\ka^2\]+\Delta A_\ka~,
\eeqa
\normalsize
with new contributions due to non-vanishing $\Delta G_S$
\beqa
\Delta A_\la&=&-d\f{F_\phi}{16\pi^2}\[3(\la^T_P)^2+2(\la^D_P)^2\]~,\\
\Delta A_\ka&=&-3d\f{ F_\phi}{16\pi^2}\[3(\la^T_P)^2+2(\la^D_P)^2\]~.
\eeqa
The soft SUSY breaking parameters for the scalars can be parameterized by 
\beqa
m_{soft}^2=\delta_m+\delta_{d}+\delta_{I}~,
\eeqa
with each part given as follows
\bit
\item The pure modulus contribution part
\beqa
\delta^m_{\tl{Q}_{L;3}}&=&m_{Q_{L;3}}M_0^2\\
&-&\f{M_0^2}{8\pi^2}\ln\(\f{M_{GUT}}{M_{mess}}\)\left\{ (q^2_{y_t}+q_{y_t})y_t^2+(q^2_{y_b}+q_{y_b})y_b^2-\f{16}{3}g_3^2-3g_2^2-\f{1}{15}g_1^2\right.\nn\\
&+&\left.\f{1}{8\pi^2}\[-y_t^2q_{y_t} K_{y_t}-y_b^2q_{y_b} K_{y_b}\]+\f{1}{8\pi^2}\[\f{8}{3} b_3^\pr g_3^2+ \f{3}{2}b_2^\pr g_2^2+\f{1}{30}b_1^\pr g_1^2\]
\ln\(\f{GUT}{M_{mess}}\) \right\},\nn\\
\delta^m_{\tl{U}^c_{L;3}}&=&m_{U^c_{L;3}}M_0^2-\f{M_0^2}{8\pi^2}\ln\(\f{M_{GUT}}{M_{mess}}\)\left\{ 2(q^2_{y_t}+q_{y_t})y_t^2-\f{16}{3} g_3^2-\f{16}{15} g_1^2\right.\\
&+&\left.\f{1}{8\pi^2}\[-2y_t^2q_{y_t} K_{y_t}\]+\f{1}{8\pi^2}\[\f{8}{3} b_3^\pr g_3^2+\f{8}{15}b_1^\pr g_1^2\]
\ln\(\f{GUT}{M_{mess}}\) \right\},\nn\\
\delta^m_{\tl{D}^c_{L;3}}&=&m_{D^c_{L;3}}M_0^2-\f{M_0^2}{8\pi^2}\ln\(\f{M_{GUT}}{M_{mess}}\)\left\{ 2(q^2_{y_b}+q_{y_b})y_b^2-\f{16}{3}g_3^2-\f{4}{15}g_1^2\right.\\
&+&\left.\f{1}{8\pi^2}\[-2y_b^2q_{y_b} K_{y_b}\]+\f{1}{8\pi^2}\[\f{8}{3} b_3^\pr g_3^2+\f{2}{15}b_1^\pr g_1^2\]
\ln\(\f{GUT}{M_{mess}}\) \right\},\nn\\
\delta^m_{\tl{L}_{L;a}}&=&m_{L_{L;a}}M_0^2\\
&-&\f{M_0^2}{8\pi^2}\ln\(\f{M_{GUT}}{M_{mess}}\)\left\{-3g_2^2-\f{3}{5}g_1^2+\f{1}{8\pi^2}\[\f{3}{2} b_2^\pr g_2^2+\f{3}{10}b_1^\pr g_1^2\]
\ln\(\f{GUT}{M_{mess}}\) \right\},\nn\\
\delta^m_{\tl{E}^c_{L;a}}&=&m_{E^c_{L;a}}M_0^2\\
&-&\f{M_0^2}{8\pi^2}\ln\(\f{M_{GUT}}{M_{mess}}\)\left\{-\f{12}{5}g_1^2+\f{1}{8\pi^2}\(\f{6}{5}b_1^\pr g_1^2\)
\ln\(\f{GUT}{M_{mess}}\) \right\},\nn\\
\delta^m_{\tl{H}_{u}}&=&m_{H_{u}}M_0^2\\
&-&\f{M_0^2}{8\pi^2}\ln\(\f{M_{GUT}}{M_{mess}}\)\left\{ 3(q^2_{y_t}+q_{y_t})y_t^2+(q^2_{\la}+q_{\la})\la^2-3g_2^2-\f{3}{5}g_1^2\right.\nn\\
&+&\left.\f{1}{8\pi^2}\[-3y_t^2q_{y_t} K_{y_t}-\la^2q_{\la} K_{\la}\]+\f{1}{8\pi^2}\[\f{3}{2}b_2^\pr g_2^2+\f{3}{10}b_1^\pr g_1^2\]
\ln\(\f{GUT}{M_{mess}}\) \right\},\nn\\
\delta^m_{\tl{H}_{d}}&=&m_{H_d}M_0^2-\f{M_0^2}{8\pi^2}\ln\(\f{M_{GUT}}{M_{mess}}\)\left\{ 3(q^2_{y_b}+q_{y_b})y_b^2+(q^2_{\la}+q_{\la})\la^2-3g_2^2-\f{3}{5}g_1^2\right.\\
&+&\left.\f{1}{8\pi^2}\[-3y_b^2 q_{y_b}K_{y_b}-\la^2q_{\la} K_{\la}\]+\f{1}{8\pi^2}\[\f{3}{2}b_2^\pr g_2^2+\f{3}{10}b_1^\pr g_1^2\]
\ln\(\f{GUT}{M_{mess}}\) \right\},\nn\\
\delta^m_{S}&=&m_{S}M_0^2-\f{M_0^2}{8\pi^2}\ln\(\f{M_{GUT}}{M_{mess}}\)\left\{\f{}{} 2(q^2_{\la}+q_{\la})\la^2+ 2(q^2_{\ka}+q_{\ka})\ka^2\right.\\
&+&\left.\f{1}{8\pi^2}\[-2\la^2q_{\la}K_{\la}-2\ka^2q_{\ka} K_{\ka}\] \right\},\nn
\eeqa
with $b^\pr_i=b_i+\Delta b_i$ being the beta function upon the messenger thresholds and
\beqa
K_{y_t}&=&\[6y_t^2 q_{y_t}+y_b^2 {q_{y_b}}+\la^2 {q_{\la}}-(\f{16}{3}g_3^2+3g_2^2+\f{13}{15}g_1^2)\]\ln\(\f{GUT}{M_{mess}}\)~,\\
K_{y_b}&=& \[y_t^2 q_{y_t}+6y_b^2 {q_{y_b}}+\la^2 {q_{\la}}-(\f{16}{3}g_3^2+3g_2^2+\f{7}{15}g_1^2)\]\ln\(\f{GUT}{M_{mess}}\)~,\\
K_{\la}&=&\[ 3y_t^2 q_{y_t}+3y_b^2 {q_{y_b}}+4\la^2 {q_{\la}}+2\ka^2 {q_{\ka}}-(3g_2^2+\f{3}{5}g_1^2)\]\ln\(\f{GUT}{M_{mess}}\)~,\\
K_{\ka}&=& \[6\la^2 {q_{\la}}+6\ka^2 {q_{\ka}}\]\ln\(\f{GUT}{M_{mess}}\).
\eeqa
Expressions for the first two generations can be obtained by setting $y_t=y_b=0$.
\item The deflected anomaly mediation part
\beqa
\delta^d_{\tl{Q}_{L;1,2}}&=&\f{F_\phi^2}{16\pi^2}\[\f{8}{3} G_3 \al^2_3+\f{3}{2}G_2\al^2_2+\f{1}{30}G_1\al^2_1\]~,\\
\delta^d_{\tl{U}^c_{L;1,2}}&=&\f{F_\phi^2}{16\pi^2}\[\f{8}{3} G_3 \al^2_3+\f{8}{15}G_1\al^2_1\]~,\\
{\delta^d_{\tl{D}^c_{L;1,2,3}}}&=&\f{F_\phi^2}{16\pi^2}\[\f{8}{3} G_3 \al^2_3+\f{2}{15}G_1\al^2_1\]~,\\
{\delta^d_{\tl{L}_{L;1,2,3}}}&=&\f{F_\phi^2}{16\pi^2}\[\f{3}{2}G_2\al_2^2+\f{3}{10}G_1\al_1^2\]~,\\
{\delta^d_{\tl{E}_{L;1,2,3}^c}}&=&\f{F_\phi^2}{16\pi^2}\f{6}{5}G_1\al_1^2~,\\
\delta^d_{{H}_u}~~~&=&\f{F_\phi^2}{16\pi^2}\[\f{3}{2}G_2\al^2_2+\f{3}{10}G_1\al^2_1\]+\f{F_\phi^2}{(16\pi^2)^2}\[\la^2G_\la+ 3y_t^2G_{y_t}\]\nn\\
&& -2 d\f{F_\phi^2}{(16\pi^2)^2}\la^2\[3(\la_P^T)^2+2(\la_P^D)^2\],\\
\delta^d_{{H}_d}~~~&=&\f{F_\phi^2}{16\pi^2}\[\f{3}{2}G_2\al^2_2+\f{3}{10}G_1\al^2_1\]+
\f{F_\phi^2}{(16\pi^2)^2}\[\la^2 G_\la+3y_b^2 G_{y_b}\]\nn\\
&&-2 d\f{F_\phi^2}{(16\pi^2)^2}\la^2\[3(\la_P^T)^2+2(\la_P^D)^2\],~
\eeqa
with
\beqa
G_{y_t}&=&6y_t^2+y_b^2+\la^2-\f{16}{3}g_3^2-3g_2^2-\f{13}{15}g_1^2,\nn\\
G_{y_b}&=&y_t^2+6y_b^2+\la^2-\f{16}{3}g_3^2-3g_2^2-\f{7}{15}g_1^2,\nn\\
G_{\la}&=&4\la^2+2\ka^2+3y_t^2+3y_b^2-3g_2^2-\f{3}{5}g_1^2~,\nn\\
G_{\ka}&=&6\la^2+6\ka^2~,
\eeqa
and $N=2$ in our scenario for
\beqa
G_i&=&Nd^2+2N d-b_i~,\\
(b_1,b_2,b_3)&=&(\f{33}{5},1,-3)~.
\eeqa
 For the third generation $\tl{Q}_{L,3},\tl{U}_L^c$, we need to include the $'y_t,y_b'$ Yukawa contributions
\beqa
\delta^d_{\tl{Q}_{L,3}}&=&\delta^d_{\tl{Q}_{L;1,2}}+F_\phi^2\f{1}{(16\pi^2)^2}\[y_t^2G_{y_t}+y_b^2G_{y_b}\]~,\\
\delta^d_{\tl{U}^c_{L,3}}&=&\delta^d_{\tl{U}^c_{L;1,2}}+F_\phi^2\f{1}{(16\pi^2)^2}2y_t^2G_{y_t}~,\\
\delta^d_{\tl{D}^c_{L,3}}&=&\delta^d_{\tl{D}^c_{L;1,2}}+F_\phi^2\f{1}{(16\pi^2)^2}2y_b^2G_{y_b}~.
\eeqa
The contributions to $\delta^I_{\tl{S}}$ can be divided into three parts
\beqa
\delta^d_{\tl{S}}=\Delta_P^A+\Delta_P^G+\Delta_P^I.
\eeqa
 We have the pure anomaly mediation part
\small
\beqa
\Delta_P^A&=&\f{F_\phi^2}{(16\pi^2)^2}\[2\la^2 G_\la
+2\ka^2 G_\ka\].
\eeqa
\normalsize
Besides, the $m^2_{\tl{S}}$ term receives new contributions involving $\la_P$ because $G_S$ is not continuous across the messenger threshold
\beqa
\Delta G_S=-\f{1}{8\pi^2}\[3(\la^T_P)^2Z_{X X}^{-1}Z_{\bar{X}\bar{X}}^{-1}+2(\la^D_P)^2 Z_{Y Y}^{-1}Z_{\bar{Y}\bar{Y}}^{-1}\].
\eeqa
So the Yukawa mediation contribution is
\small
\beqa
\Delta_P^G=-\f{d^2F_\phi^2}{4(8\pi^2)}\[3(\la_P^T)^2 \(G_{{\la_P^T}}^+\) + 2(\la_P^D)^2 \(G_{{\la_P^D}}^+\)\]
+\f{d^2F_\phi^2}{16\pi^2}\(\la^2\Delta G_{{\la}}+\ka^2\Delta G_{{\ka}}\),
\eeqa
\normalsize
with
\small
\beqa
G_{{\la_P^T}}^+&=&-\f{1}{8\pi^2}\[5(\la_P^T)^2+2(\la_P^D)^2+2\la^2+2\ka^2+2(\la_X^T)^2-\(\f{16}{3}g_3^2+\f{4}{15}g_1^2\)\],\\
G_{{\la_P^D}}^+&=&-\f{1}{8\pi^2}\[3(\la_P^T)^2+4(\la_P^D)^2+2\la^2+2\ka^2+2(\la_X^D)^2-\(3g_2^2+\f{3}{5}g_1^2\)\],\\
\Delta G_{{\ka}}&=&-\f{1}{8\pi^2}3\[3(\la_P^T)^2+2(\la_P^D)^2\],\\
\Delta G_{{\la}}&=&-\f{1}{8\pi^2}\[3(\la_P^T)^2+2(\la_P^D)^2\].
\eeqa
\normalsize
The anomaly-gauge(Yukawa) mixing term is given by
\beqa
\Delta_P^I=-\f{2 d F^2_\phi}{(16\pi^2)^2}\left\{2\la^2\[3(\la_P^T)^2
+2(\la_P^D)^2\]+6\ka^2\[3(\la_P^T)^2+2(\la_P^D)^2\]\f{}{}\right\}.
\eeqa
\item The interference terms involving the Kahler modulus $'T'$:
\beqa
\delta^I_{\tl{Q}_L}&=&\f{M_0 F_\phi}{8\pi^2}\[y_t^2\(q_{y_t}-\f{1}{8\pi^2}K_{y_t}\)+y_b^2\(q_{y_b}-\f{1}{8\pi^2}K_{y_b}\)\right.\nn\\
&&\left.-\(\f{8}{3}\f{g_3^4}{g_3^2(GUT)}+\f{3}{2}\f{g_2^4}{g_2^2(GUT)}+\f{1}{30}\f{g_1^4}{g_1^2(GUT)}\)\]~,\\
\delta^I_{\tl{U}^c_L}&=&\f{M_0 F_\phi}{8\pi^2}\[2y_t^2\(q_{y_t}-\f{1}{8\pi^2}K_{y_t}\)-\(\f{8}{3}\f{g_3^4}{g_3^2(GUT)}+\f{8}{15}\f{g_1^4}{g_1^2(GUT)}\)\]~,\\
\delta^I_{\tl{D}^c_L}&=&\f{M_0 F_\phi}{8\pi^2}\[2y_b^2\(q_{y_b}-\f{1}{8\pi^2}K_{y_b}\)-\(\f{8}{3}\f{g_3^4}{g_3^2(GUT)}+\f{2}{15}\f{g_1^4}{g_1^2(GUT)}\)\]~,\\
\delta^I_{\tl{L}_L}&=&-\f{M_0 F_\phi}{8\pi^2}\(\f{3}{2}\f{g_2^4}{g_2^2(GUT)}+\f{3}{10}\f{g_1^4}{g_1^2(GUT)}\)~,\\
\delta^I_{\tl{E}_L^c}&=&-\f{M_0 F_\phi}{8\pi^2}\(\f{6}{5}\f{g_1^4}{g_1^2(GUT)}\)~,\\
\delta^I_{H_u}&=&\f{M_0 F_\phi}{8\pi^2}\[3y_t^2\(q_{y_t}-\f{1}{8\pi^2}K_{y_t}\)+\la^2\(q_{\la}-\f{1}{8\pi^2}K_{\la}\)\right.\nn\\
&& \left.-\(\f{3}{2}\f{g_2^4}{g_2^2(GUT)}+\f{3}{10}\f{g_1^4}{g_1^2(GUT)}\)\]~,\\
\delta^I_{H_d}&=&\f{M_0 F_\phi}{8\pi^2}\[3y_b^2\(q_{y_b}-\f{1}{8\pi^2}K_{y_b}\)+\la^2\(q_{\la}-\f{1}{8\pi^2}K_{\la}\)\right.\nn\\
&&\left.-\(\f{3}{2}\f{g_2^4}{g_2^2(GUT)}+\f{3}{10}\f{g_1^4}{g_1^2(GUT)}\)\]~,\\
\delta^I_{\tl{S}}&=&\f{M_0 F_\phi}{8\pi^2}\[2\la^2\(q_{\la}-\f{1}{8\pi^2}K_{\la}\)+2\ka^2 \(q_{\ka}-\f{1}{8\pi^2}K_{\ka}\)\]+\Delta_P^{TX}(m_{\tl{S}^2}) ~.~
\eeqa
Note that the expressions for sfermions are hold for the third generation, the first two generation can be obtained by setting $y_t,y_b\ra0$.
Within the expressions, modular weight $l_i=1$ for gauge couplings are used.

The previous contributions are just the anomaly-modulus interference part. Possible modulus-gauge interference part will also appear in our scenario. The anomalous dimensions for all fields except $S$ are continuous across the messenger threshold,
so their modulus-gauge interference contributions vanish.
As noted previously, the anomalous dimension for $S$ is discontinuous across the messenger threshold, so we have the new $T,X$ interference contributions to $m^2_{\tl{S}}$
\beqa
\Delta_P^{TX}(m_{\tl{S}^2})&=&-\f{ d M_0 F_\phi}{8\pi^2}\[3(\la_P^T)^2\(q_{\la_P^T}-\f{1}{8\pi^2}K_{\la_P^T}\)+2(\la_P^D)^2\(q_{\la_P^D}-\f{1}{8\pi^2}K_{\la_P^D}\)\],\nn\\
\label{interference:g-m}
\eeqa
with
\small
\beqa
K_{\la_P^T}&=&\[5(\la_P^T)^2q_{\la_P^T}+2(\la_P^D)^2q_{\la_P^D}+2\la^2{q_{\la}}+2\ka^2{q_{\ka}}+2(\la_X^T)^2{q_{\la_X^T}}-\(\f{16}{3}g_3^2+\f{4}{15}g_1^2\)\]
\ln\(\f{GUT}{M_{mess}}\)
~,\nn\\
K_{\la_P^D}&=& \[3(\la_P^T)^2q_{\la_P^T}+4(\la_P^D)^2q_{\la_P^D}+2\la^2{q_{\la}}+2\ka^2{q_{\ka}}+2(\la_X^D)^2{q_{\la_X^D}}-\(3g_2^2+\f{3}{5}g_1^2\)\]\ln\(\f{GUT}{M_{mess}}\)
~.\nn\\
\eeqa
\normalsize
Here
\beqa
q_{\la_P^T}&=& m_{S}+m_{\tl{X}_1}+m_{X_2}~,\nn\\
q_{\la_P^D}&=& m_{S}+m_{\tl{Y}_1}+m_{Y_2}~,\nn\\
q_{\la_X^T}&=& m_{X}+m_{\tl{X}_m}+m_{X_m}~,\nn\\
q_{\la_X^D}&=& m_{X}+m_{\tl{Y}_m}+m_{Y_m}~.
\eeqa

\eit

\section{\label{sec-4} Numerical Results}

After fixing the modular weights, the remaining free parameters in our scenario are
  \beqa
  d, \al, M_{mess}, M_0, \la,\ka,\la_P^D,\la_P^T,\la_X^D,\la_X^T
  \eeqa
  with $F_\phi/(16\pi^2)\approx \al M_0$ and the simplest choice $\la_P^D=\la_P^T=\la_X^D=\la_X^T=\la_0$
in our numerical study. Note that for later convenience, the definition of $\al$ is four times smaller
than $\al^\pr$ that appears in eq.(\ref{alpha}).
The ratio $\al$ between $F_\phi/(16\pi^2)$ and $M_0$ holds in the messenger scale and in general
is different from its value at the GUT scale. We choose a positive $\al$ in our numerical study.
For a negative $\al$, virtual mirage unification at a super-GUT energy scale will appear.

We need to check if successful EWSB conditions are indeed fulfilled.
In fact, the soft SUSY mass $m_{H_u}^2,m_{H_d}^2,m_S^2$ can be reformulated into $\mu,\tan\beta,M_Z^2$
by the minimum conditions of the scalar potential. Usually, $M_A$ can be used to replace $A_\ka$
  \beqa
  M_A^2=\f{2\mu_{eff}}{\sin2\beta}B_{eff}~, ~~\mu_{eff}\equiv\la\langle s\rangle~,~~~B_{eff}=(A_\la+\ka \langle s\rangle).
  \eeqa
In order to transform $m_{H_u}^2,m_{H_d}^2,m_S^2$ into $\mu,\tan\beta,M_Z^2$, we use the following approximation
\beqa
|\mu_{eff}|^2&=&-\f{M_Z^2}{2}-m_{H_u}^2+\f{1}{\tan^2\beta}(m_{H_d}^2-m_{H_u}^2)+{\cal O}(1/\tan^4\beta)~,\\
\sin2\beta&=&\f{2B_{eff}\mu_{eff}}{m_{H_u}^2+m_{H_d}^2+2|\mu_{eff}|^2+\la^2v^2}~,
\eeqa
 to obtain $\mu$ and $\tan\beta$ iteratively.

   We use NMSSMTools5.2.0\cite{NMSSMTools} to scan the whole parameter space. The parameters are chosen to satisfy:
 \beqa
&&10^{15} {\rm GeV} >M_{mess}> 10^{5} {\rm GeV}~,~~~ 100 {\rm TeV}>  M_0 > 0.1{\rm TeV}~,\\
&&16 >\al>0~,~~~5>d>0~,~~~0.7>\la,\ka>0~,~\sqrt{4\pi}>\la_0 >0~.~~
 \eeqa

 In our scan, we impose the following constraints:
\bit
\item[(I)]  The conservative lower bounds on SUSY particles \cite{atlas:gluino,atlas:stop} from the LHC
    \bit
    \item  Gluino mass: $m_{\tl{g}} \gtrsim 1.8$ TeV .
    \item  Light stop mass: $m_{\tl{t}_1} \gtrsim 0.85$ TeV .
    \item  Light sbottom mass $m_{\tl{b}_1} \gtrsim 0.84$ TeV.
    \item  Degenerated first two generation squarks $m_{\tl{q}} \gtrsim 1.0 \sim 1.4$ TeV.
    \eit
\item[(II)] The CP-even component $S_2$ in the Goldstone-$'eaten'$ combination of $H_u$ and $H_d$ doublets corresponds to the SM Higgs.
 The $S_2$ dominated CP-even scalar should lie in the combined mass range for the Higgs boson: $122 {\rm ~GeV}<M_h <128 {\rm ~GeV}$
from ATLAS and CMS data \cite{ATLAS:higgs,CMS:higgs}. Note that the uncertainty is 3 GeV instead of the default 2 GeV
because a large $\lambda$ may induce an additional 1 GeV correction to $m_h$ at two-loop level \cite{NMSSM:higgs2loop},
which is not included in the NMSSMTools.

\item[(III)] The relic density of the neutralino dark matter should satisfy the Planck data
$\Omega_{DM} = 0.1199\pm 0.0027$ \cite{Planck}
in combination with the WMAP data \cite{WMAP} (with a $10\%$ theoretical uncertainty).

\item[(IV)] The electroweak precision observables \cite{precision} and the lower bounds on neutralinos
and charginos, including the invisible decay bounds for $Z$-boson. The most stringent constraints of
LEP require $m_{\tl{\chi}^\pm}> 103.5 {\rm GeV}$ and
     the invisible decay width $\Gamma(Z\ra \tl{\chi}_0\tl{\chi}_0)<1.71~{\rm MeV}$,
     which is consistent with the $2\sigma$ precision EW measurement $\Gamma^{non-SM}_{inv}< 2.0~{\rm MeV}$ \cite{LEP}.

\item[(V)] Flavor constraints \cite{B-physics} from B-meson rare decays
  \beqa
  &&1.7\tm 10^{-9} < Br(B_s\ra \mu^+ \mu^-) < 4.5\tm 10^{-9}~,\\
  && 0.85\tm 10^{-4} < Br(B^+\ra \tau^+\nu) < 2.89\tm 10^{-4}~,\\
  && 2.99\tm 10^{-4} < Br(B_S\ra X_s \gamma) < 3.87\tm 10^{-4}~.
   \eeqa

\item[(VI)] The tension between the theoretical prediction and the experimental value for
the muon anomalous magnetic moment should be ameliorated by additional positive SUSY contributions.
The E821 experimental result for the muon $g-2$ at the Brookhaven AGS \cite{Mg-2:collaboration} was given by
 \beqa
a_\mu^{\rm expt} =116592089(63)\times 10^{-11}~,
 \eeqa
which is larger than the SM prediction\cite{Mg-2:SM}
\beqa
a^{\rm SM}_\mu =116591834(49)\times 10^{-11}~.
\eeqa
  The deviation is about $3\sigma$
 \beqa
\Delta a_\mu({\rm expt - SM}) = (255\pm 80)\times 10^{-11}.
\eeqa
We adopt a conservative estimation $4.7\tm 10^{-10}\lesssim\Delta a_\mu\lesssim52.7\tm 10^{-10}$ in our numerical results.
\eit

We should note that the numerical results depend crucially on whether the 125 GeV Higgs is the lightest CP-even scalar (Type {\bf A}) or the second lightest CP-even scalar(Type {\bf B}). We have the following discussions
 \bit
 \item

  The low energy soft SUSY breaking spectrum of NMSSM, determined from a top-down approach by a UV-completed
theory, is always bothered by the requirement to achieve successful EWSB. As noted previously, EWSB conditions in NMSSM in general
require a large VEV for the singlet and consequently prefer a negative $m_S^2$ and/or large $A_\la,A_\ka$ for the singlet potential.
However, ordinary mirage mediation scenarios always predict large positive values for $m^2_S$ and not very large $A_\la,A_\ka$, suppressing the singlet VEV.
In our scenario, because of possible negative contributions to $m_S^2$ from new Yukawa interactions, stringent constraints from successful EWSB can be ameliorated. The observed 125 GeV Higgs mass, whether it is the lightest or the second-lightest CP-even scalar in NMSSM,
can also be successfully accommodated in our scenario.

\begin{figure}[htb]
\begin{center}
\includegraphics[width=2.9in]{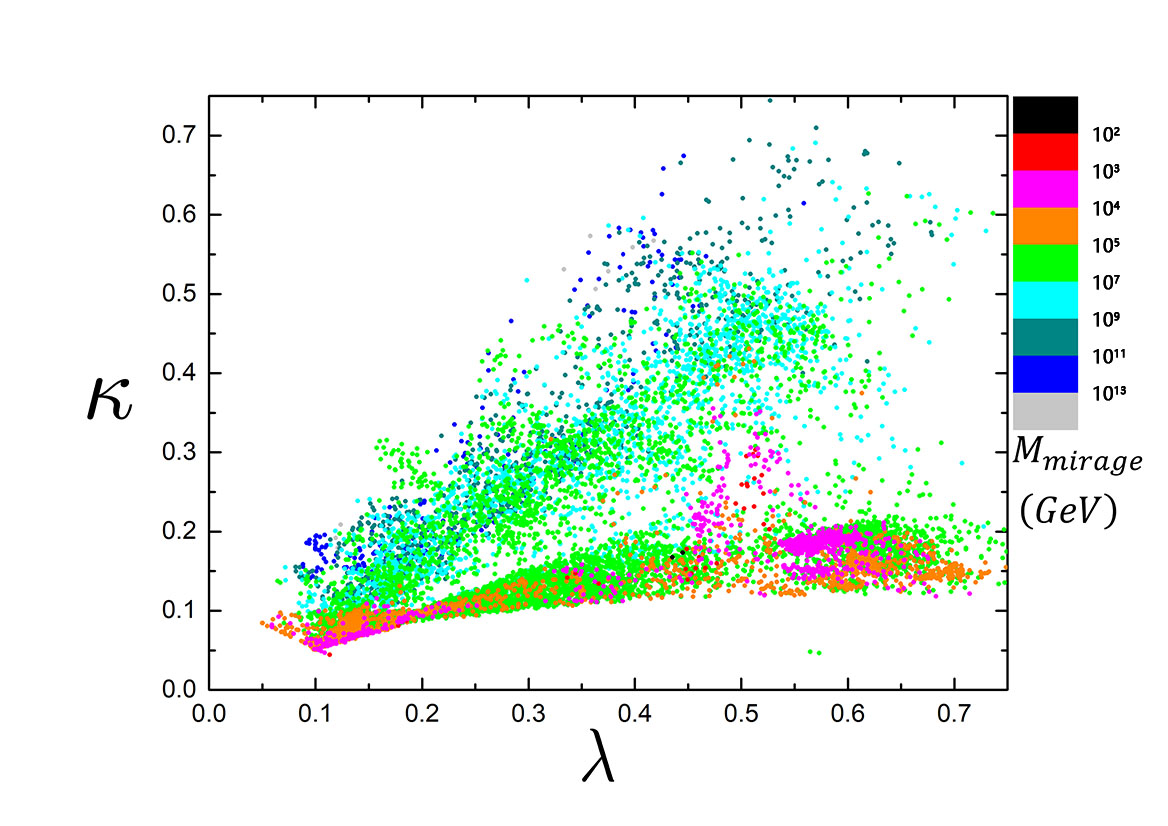}
\includegraphics[width=2.9in]{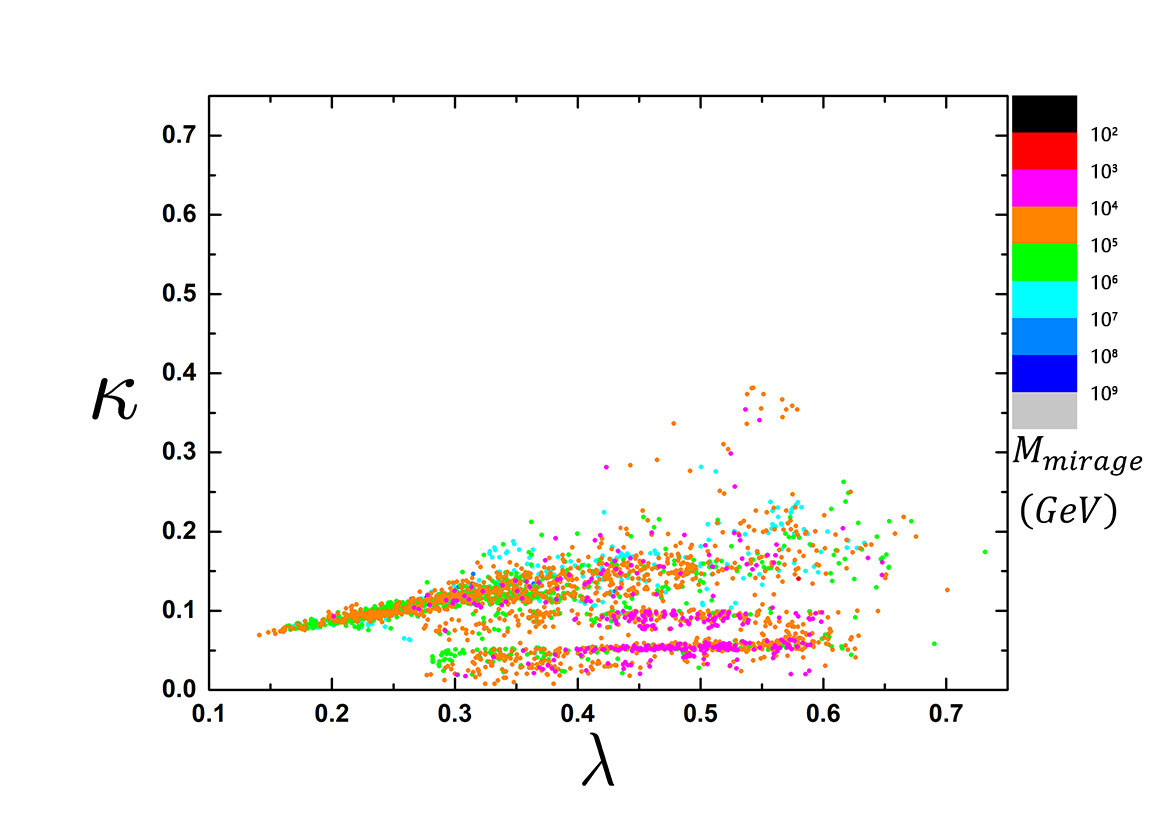}\\
\end{center}
\vspace{-.5cm}
\caption{ The values of $(\la,\ka)$ that can satisfy the EWSB conditions and at the same time accommodate
the 125 GeV Higgs boson as the lightest (left panel) or second lightest (right panel) CP-even scalar are shown. The mirage unification scales for gaugino masses are also shown in different colors.
All points satisfy the constraints  (I-V).}
\label{fig1}
\end{figure}
We can see from Fig.\ref{fig1} that many samples of ($\la,\ka$) can survive
the EWSB conditions in NMSSM.
In contrast to the numerical results of TeV mirage mediation in \cite{mirageNMSSM1} within which the allowed ($\la,\ka$) only take
values near (0.7,0.1), some portion of ($\la,\ka$) parameter space can survive all constraints in Type {\bf A}.

We should note that such a difference has the following reasons
\bit
\item  The choices of the modular weights in \cite{mirageNMSSM1}, for example, the values of $m_{Q_L^3}$ and $m_{t_L^c}$ etc, are different to ours which are given in Eqn.(\ref{modular:weight}).
\item  New ingredients, such as the messenger sector which can cause additional deflection of the RGE trajectory, will introduce new free parameters. After all, the mirage mediation scenario can be seen as a special case of our scenario with gauge couplings(Yukawa couplings) being switched off.
\item  The mirage unification scale is set to lie at TeV scale in \cite{mirageNMSSM1}. It is known that simple mirage unification for soft parameters would in general be spoiled in deflected mirage mediation scenario. However, $'mirage'$ unification for gaugino masses persists which can be proven in our Wilsonian approach (see Appendix A for details).

      In our scenario, the mirage unification scale for gaugino masses is not constrained to lie at TeV scale. However, we can see from the mirage unification scales shown by different colors in Fig.\ref{fig1} that even if such scales are required to lie at TeV scale, vast parameter space, which is larger than the numerical result of Ref.(\cite{mirageNMSSM1}), can survive the constraints from (I-V).
\eit

Note that the vacuum stability bounds are also taken into account in our numerical studies, which impose
stringent constraints on scenario in \cite{mirageNMSSM1}. In Type B in which the 125 GeV Higgs is the second
lightest CP-even scalar, the allowed ($\la,\ka$) parameter space is also much bigger
than that in \cite{mirageNMSSM1}.  The Higgs mass can be increased by 8 GeV through
the mixing with the singlet component for large $\tan\beta$ and $\lambda\lesssim 0.04$.

\begin{figure}[htb]
\begin{center}
\includegraphics[width=2.9in]{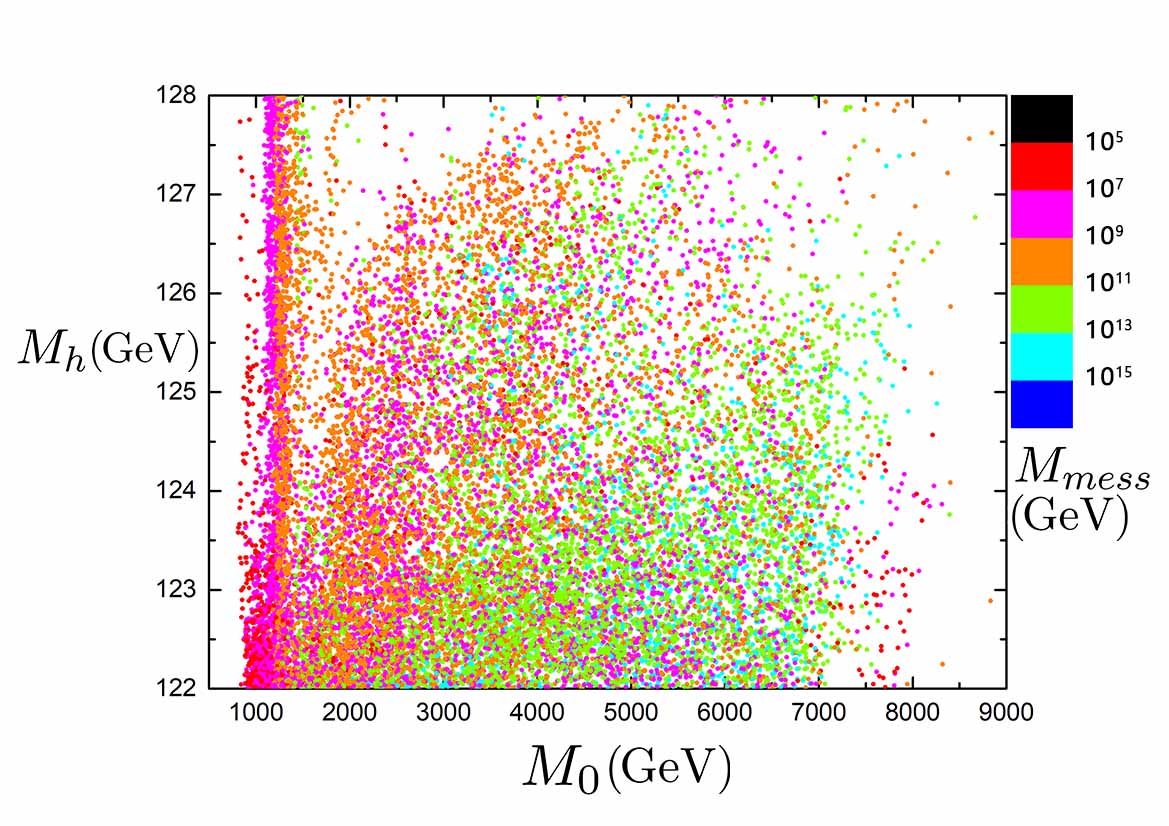}
\includegraphics[width=2.9in]{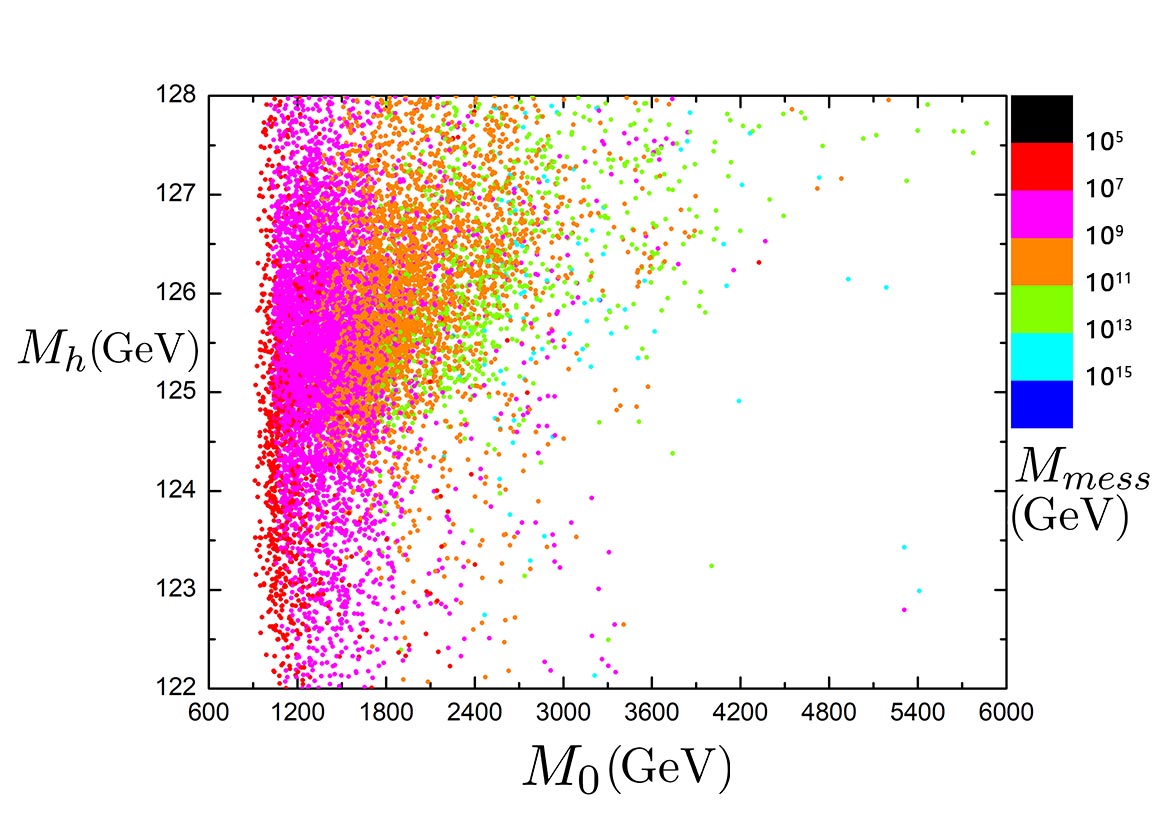}\\
\includegraphics[width=2.9in]{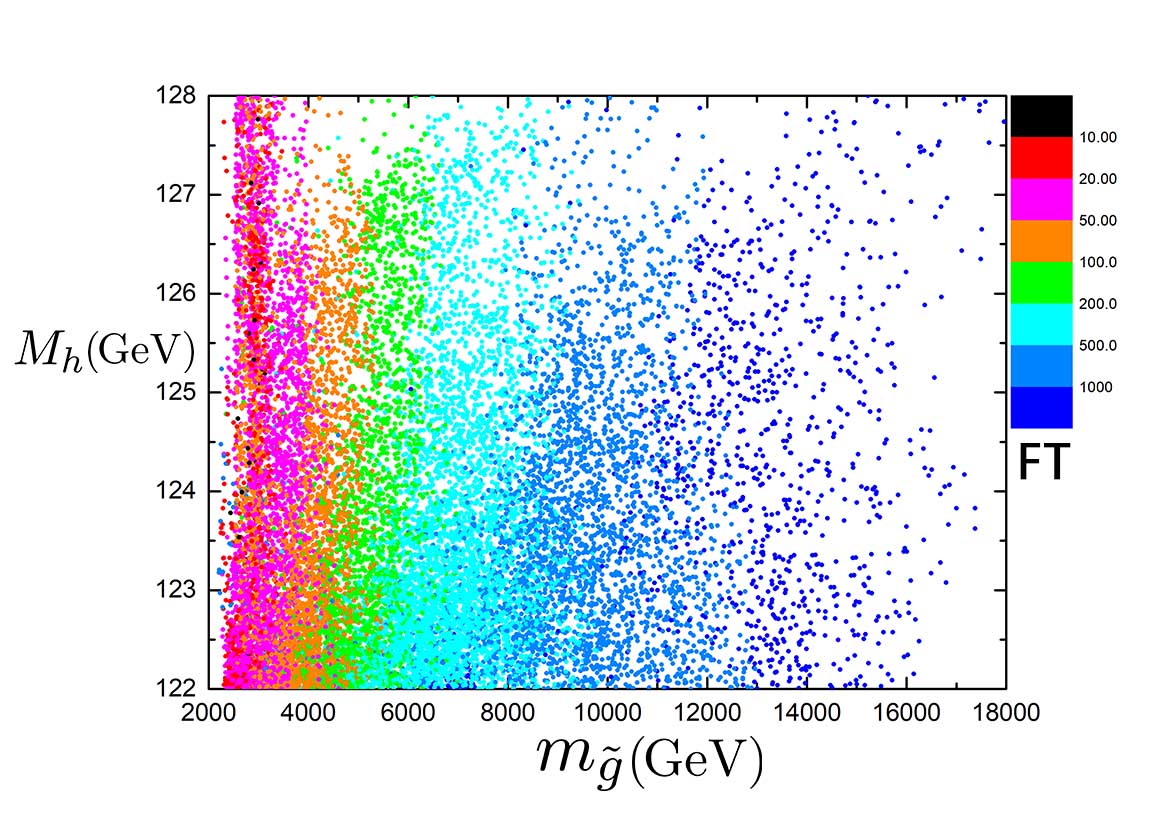}
\includegraphics[width=2.9in]{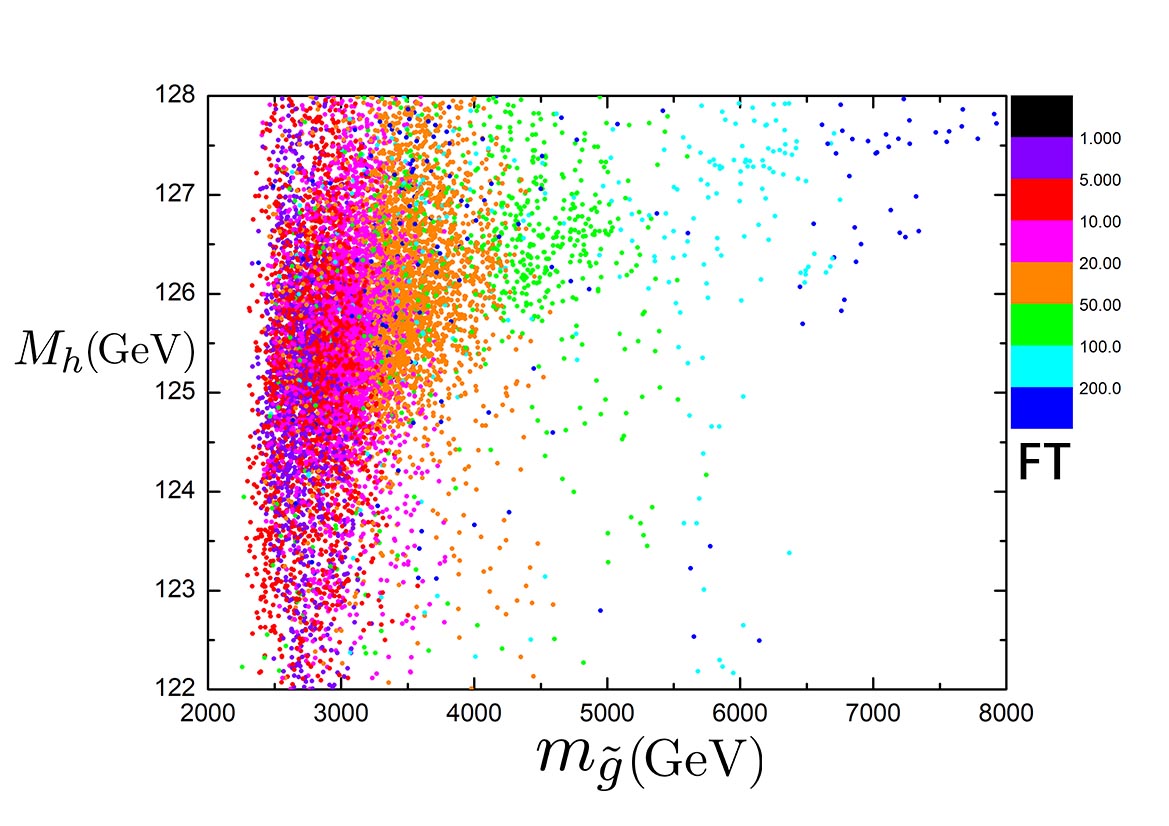}\\
\includegraphics[width=2.9in]{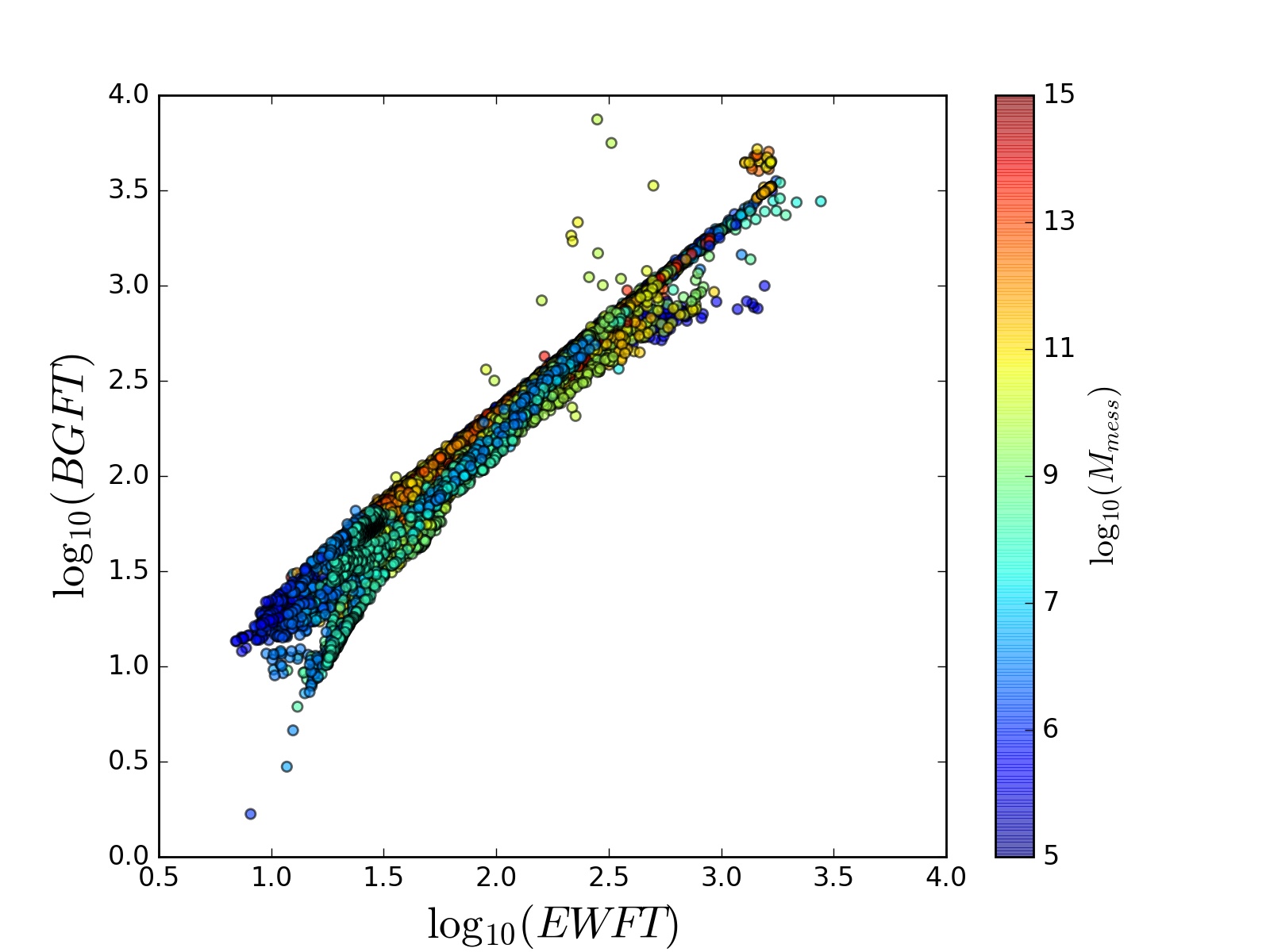}
\includegraphics[width=2.9in]{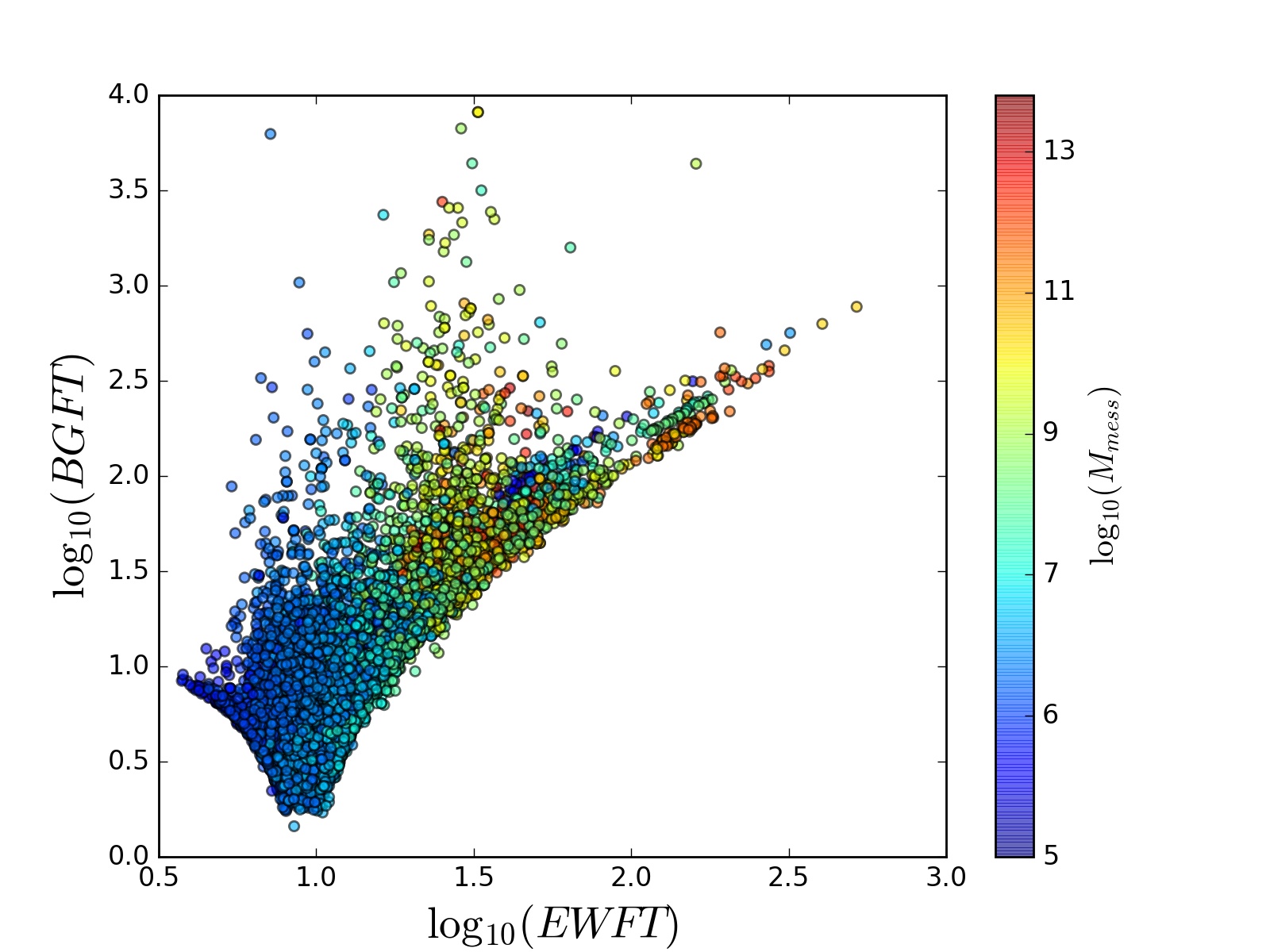}\\
\end{center}
\vspace{-.5cm}
\caption{ Scatter plots of the SM-like Higgs mass $m_h$ versus the modulus mediation parameter $M_0$
and gluino mass $m_{\tl{g}}$ for Type A (left panels) or Type B (right panels).
The fine-tuning measures given in the middle of the panels ($m_h$ vs $m_{\tl{g}}$) are the
Barbieri-Giudice fine tuning (BGFT) measures.  The comparisons between the BGFT measure versus
the electroweak fine tuning(EWFT) measure for Type A and Type B are shown in the last two panels,
respectively. All samples satisfy the constraints (I-V).}
\label{fig2}
\end{figure}

It can be seen from Fig.\ref{fig2} that the modulus mediation contribution $M_0$ is bounded
to lie between 1 TeV to 8.5 TeV for Type {\bf A}.
A small $M_0$ always prefers a low messenger scale $M_{mess}$. A large $M_0$, which controls the whole soft SUSY breaking parameters
to be heavy, can easily accommodate the SM-like Higgs mass because of large loop corrections from heavy stops
in addition to the tree-level contributions involving $\la$.
The value of $M_0$ is upper bounded to be less than about 5.5 TeV for Type {\bf B}, which sets an upper bound for the soft SUSY breaking parameters, especially for the gluino masses.
 The gluino mass, which is determined by the scale of $M_0$, is bounded to below 16 TeV for Type {\bf A} and below 8 TeV for Type {\bf B}, respectively.

\item
The Barbieri-Giudice fine-tuning(BGFT) \cite{BGFT} measure is defined as
          \beqa
          \Delta_{BG}\equiv \max\limits_{i}\left|\f{\pa \ln M_Z^2}{\pa \ln a_i}\right|~,
          \eeqa
where $'a_i'$ stands for the set of parameters defined at the input scale.

There are two mass scales for the soft SUSY parameters in our scenario,
one is the scale that characterize the modulus contributions $M_0$ and the other
is the scale that characterize the anomaly contribution $F_\phi$.
The latter one is rewritten into a dimensionless quantity $\al$ by eqn(\ref{al:def}).

   We can also calculate the electroweak fine-tuning(EWFT) measure $\Delta_{EW}$
    of the survived points defined in \cite{radiative:natural}
   \beqa
   \Delta_{EW}\equiv \max\limits_{i}(C_i)/\(\f{m_Z^2}{2}\),
   \eeqa
with the relevant terms (see \cite{radiative:natural})
\beqa
C_{H_u}&=&\left|-\f{m^2_{H_u}\tan^2\beta}{\tan^2\beta-1}\right|~,~
C_{H_d}=\left|\f{m^2_{H_d}}{\tan^2\beta-1}\right|~,~
C_{\mu}=\left|-\mu_{eff}^2\right|~,~\nn\\
C_{\Sigma_{u}^u(\tl{t}_{1,2})}&=&
\f{\tan^2\beta}{\tan^2\beta-1}\left|\f{3}{16\pi^2}F(m_{\tl{t}_{1,2}}^2)\left[y_t^2-g_Z^2\mp \f{y_t^2 A_t^2-8g_Z^2(\f{1}{4}-\f{2}{3}x_w)\Delta_t}{m_{\tl{t}_{2}}^2-m_{\tl{t}_{1}}^2}\right]\right|~,\nn\\
C_{\Sigma_{d}^d(\tl{t}_{1,2})}&=&\f{1}{\tan^2\beta-1}\left|\f{3}{16\pi^2}F(m_{\tl{t}_{1,2}}^2)\left[g_Z^2\mp \f{y_t^2 \mu_{eff}^2+8g_Z^2(\f{1}{4}-\f{2}{3}x_w)\Delta_t}{m_{\tl{t}_{2}}^2-m_{\tl{t}_{1}}^2}\right]\right|~,
\eeqa
where $x_w=\sin^2\theta_W$ and
\beqa
\Delta_t&=&\f{(m_{\tl{t}_{L}}^2-m_{\tl{t}_{R}}^2)}{2}+M_Z^2\cos2\beta(\f{1}{4}-\f{2}{3}x_w)~,\nn\\
F(m^2)&=& m^2\(\log\f{m^2}{m_{\tl{t}_{1}}m_{\tl{t}_{2}}}-1\)~.
\eeqa

  We can see from Fig.\ref{fig2} that the BGFT measure in our scenario can range
from ${\cal O}(1)$ to ${\cal O}(1000)$. In fact, the lowest BGFT can reach ${\cal O}(1)$ for Type {\bf B}.
 Such a low fine-tuning possibly indicates that our scenario is natural.
 We also compare the BGFT measure of the survived points with their corresponding EWFT measure
in the last two panels of Fig.\ref{fig2}.
It can be seen that the calculated EWFT measure is positively correlated
to the corresponding BGFT measure. In most of allowed parameter space,
the EWFT and BGFT measures take values of the same order.
Besides, the EWFT measure,
which can be thought of as providing a lower bound on the electroweak fine-tuning,
is always smaller than that of the BGFT measure\cite{FT:compare}.

In some of the allowed region, the EWFT measure as well as BGFT measure is rather
low (being ${\cal O}(1)$).
As emphasized in \cite{FT:compare},  EWFT measure is a necessary, albeit not sufficient,
measure of electroweak fine-tuning. Low $\Delta_{EW}$ need not necessarily mean
the model is not fine-tuned. Rather, it may indicate the possibility that some
model might exist with low fine-tuning which might be hidden by the naive application
of $\Delta_{BG}$.

It is known that low fine-tuning needs light stops as well as a small effective $\mu$,
which are naively determined by the dimensional parameter $M_0$
that controls the whole soft SUSY spectrum with moderate values of $\al$.
The lower the $M_0$ (consequently the lower gluino mass), the lower value of the BGFT(EWFT) measure.

\begin{figure}[htb]
\begin{center}
\includegraphics[width=2.9in]{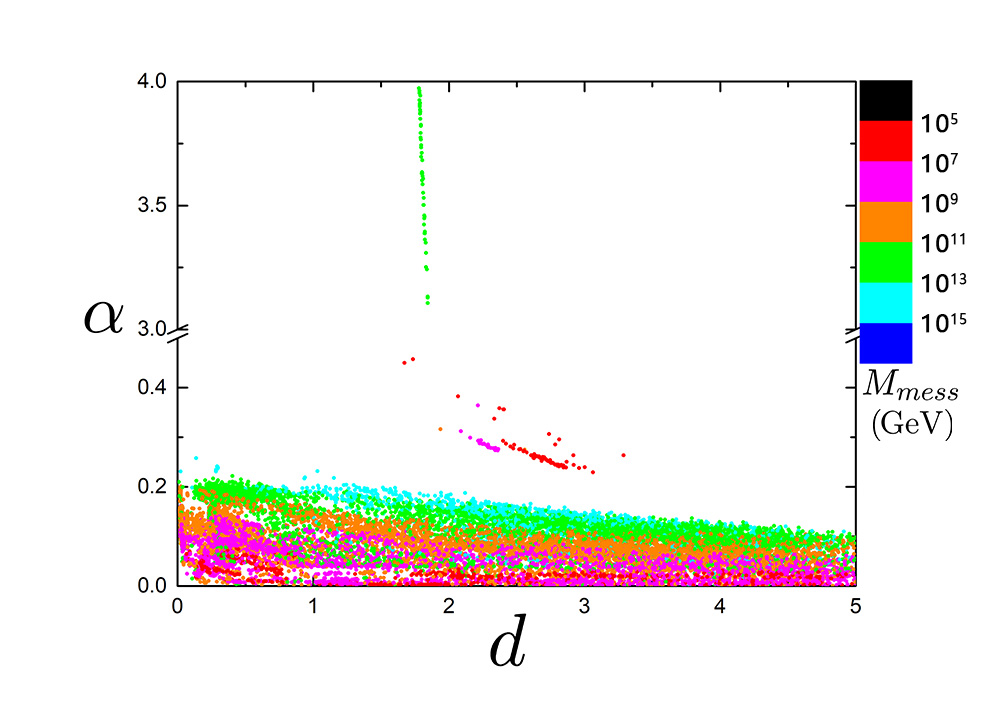}
\includegraphics[width=2.7in]{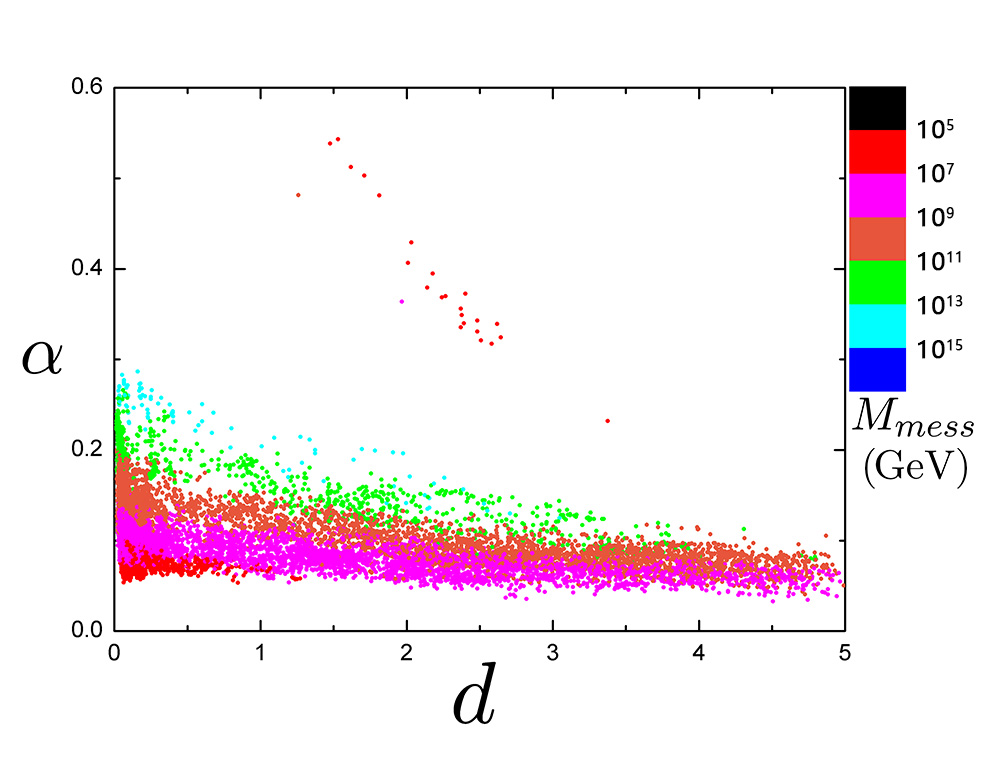}\\
\end{center}
\vspace{-.5cm}
\caption{The allowed regions for the deflection parameter $'d'$ versus $\al$,
which parametrize the relative size between the anomaly mediation and the modulus mediation.
All samples satisfy the constraints (I-V).  }
\label{fig3}
\end{figure}
\begin{figure}[htb]
\begin{center}
\includegraphics[width=2.9in]{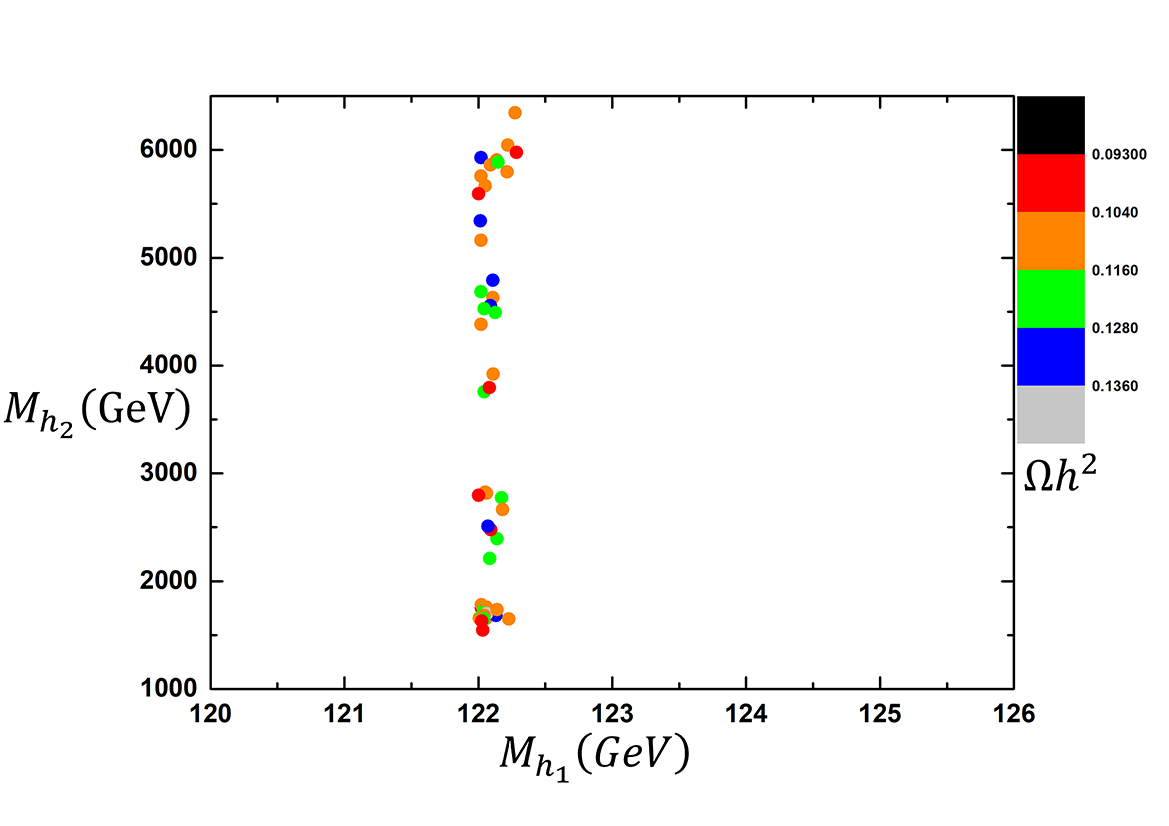}
\includegraphics[width=2.9in]{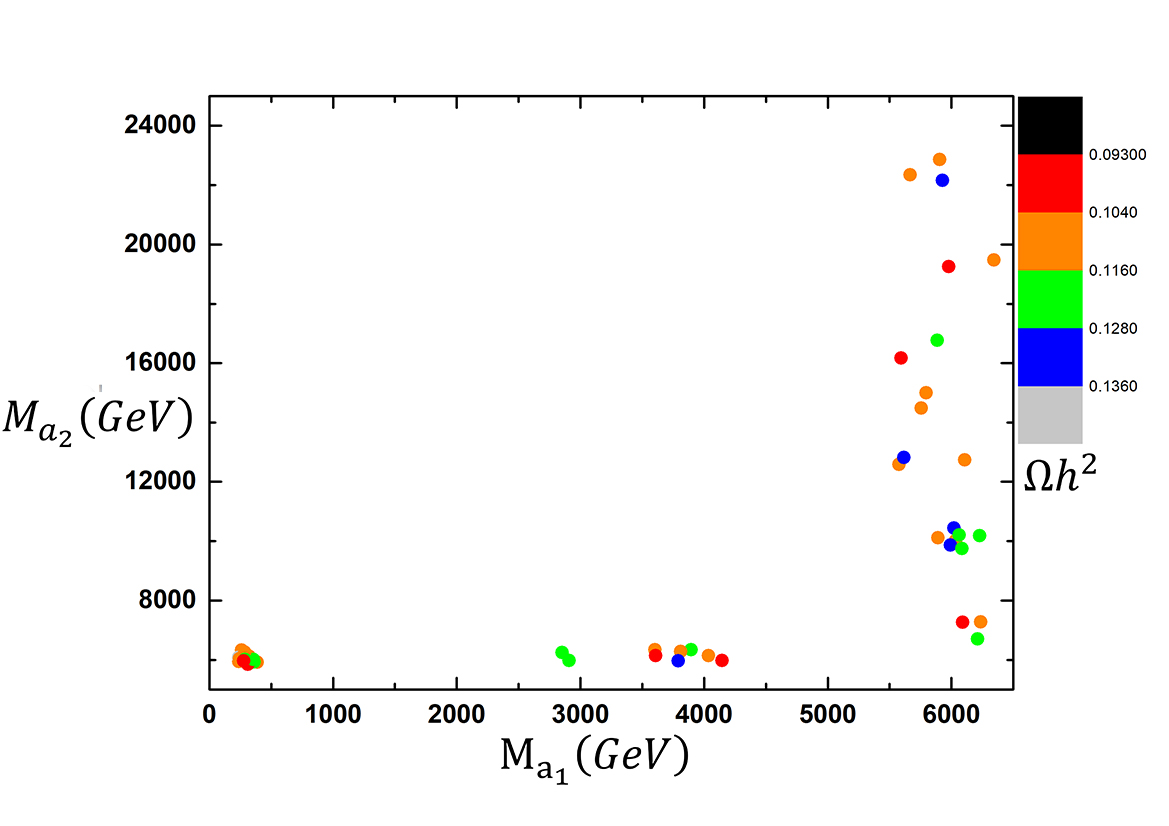}\\
\end{center}
\vspace{-.5cm}
\caption{ The masses of the Higgs scalars with $d\approx 1.8$. All samples satisfy the constraints (I-V).  }
\label{fig3:d1.8}
\end{figure}
\item A positive deflection parameter $'d'$ is always favored to solve the tachyonic slepton problem in the deflected AMSB for fewer messenger species.
In the deflected mirage mediation scenarios, if the modulus contribution is subdominant, a realistic model still prefers
a positive deflection parameter $'d'$ with less messenger species.
As the parameter $\al\equiv \al^\pr/4$ determines the relative size of the contributions
between the anomaly mediation and the modulus mediation, a large value of $\al$, which indicates
small modulus mediation contributions, needs a large positive deflection parameter $'d'$ to avoid
tachyonic sleptons. We check that large negative values of $'d'$ are mostly ruled out by the EWSB condition and tachyonic sfermions. It is obvious in the left panel of Fig.\ref{fig3} that the deflection parameter is constrained to lie at about 1.8 to tune the tachyonic slepton masses to positive values by additional gauge and Yukawa mediation
contributions in the region with a large $\al$.
It can be seen in Fig.\ref{fig3:d1.8} that the corresponding lightest CP-even Higgs mass should
lie at a very narrow band centered at about $122.1$ GeV with $d\approx 1.8$.
Besides, the second-lightest CP-odd scalar $a_2$ is constrained to lie near 6000 GeV if the
lightest CP-odd scalar $a_1$ is lighter than 5000 GeV while the lightest CP-odd scalar is
constrained to lie near 6000 GeV if the second-lightest CP-odd scalar is heavier than 6000 GeV.

 In our scenario, the quantity $'4 \al d'$ can approximately measure the relative size of deflection contributions (by gauge or Yukawa mediation) to the modulus mediation contributions. We can see from Fig.\ref{fig3} that the deflection contributions can be dominant in a large portion of the surviving parameter space.

Besides, it can also be seen from Fig.\ref{fig3} that
 in the modulus mediation dominated regions, that is small $\al$ with $d=0$,
 realistic NMSSM spectrums can be possible. This indicates that pure mirage mediation without deflection, which is a special case of our scenarios, can lead to realistic NMSSM spectrum even though it is stringently constrained by EWSB conditions and 125 GeV Higgs. This conclusion agrees with that of \cite{mirageNMSSM1}.
Additional deflection with $d\neq 0$ from messenger sector can enlarge the possible choice of $\al$ in mirage mediation, rendering the mirage-type scenarios  more natural.

\item     From eq.(\ref{gaugino}) in the appendix, we can see the gaugino ratio at the EW scale
\small
\beqa
M_3:M_2:M_1
\approx 6\cdot[\f{1}{g_3^2}+{\al}(-3-2d)]:2\cdot[\f{1}{g_2^2}+{\al}(1-2d)]:[\f{1}{g_1^2}+{\al}(6.6-2d)]
                       \label{ratio}
                       \eeqa
\normalsize
where $g_1,g_2,g_3$ take values at the GUT scale. On the other hand, the singlino mass is determined
by $\ka$ and $\langle s\rangle$, which rescales the effective $\mu_{eff}$ parameter by a factor $2\ka/\la$.
In general, a pure singlino-like LSP tends to have a too large relic density due to a comparatively
small annihilation cross section because of its small couplings to SM particles. Non-negligible higgsino contents within singlino-dominated neutralino DM can open several annihilation channels and be helpful to reduce the DM relic density to right amount.

In Type {\bf A}, the neutralino DM is either singlino-dominant or bino-dominant, each possibility contains non-negligible higgsino components.
In the bino-dominant regions, the LSP annihilate dominantly into pairs of gauge bosons ($W^+W^-,ZZ$) and (doublet-like) Higgs bosons ($W^\pm H^\pm, ZH, HA$) via s-channel Z or Higgs exchange, as well as through t-channel neutralino and chargino exchange processes.
In the singlino-dominant regions, the singlino-like
LSP (with the presence of non-negligible higgsino components) can annihilates via the t-channel $\chi^0_1$ exchange into pairs of mostly singlet-like $H_1$ and $A_1$ by enhanced $\chi_1^0 \chi_1^0 H_1(A_1)$ couplings. Co-annihilation with heavier $\chi^0_2$
(for $\Delta m\lesssim 10 {\rm GeV}$) will also efficiently reduce the singlino relic abundance
to a proper $\Omega_{DM}$. Besides, the annihilation channels $\chi_1^0\chi_1^0\ra t\bar{t},b\bar{b}$ 
can also be important. Similar DM annihilation channels exist for Type {\bf B} in which the neutralino DM is mainly singlino-like with non-negligible higgsino components.

\begin{figure}[htb]
\begin{center}
\includegraphics[width=2.9in]{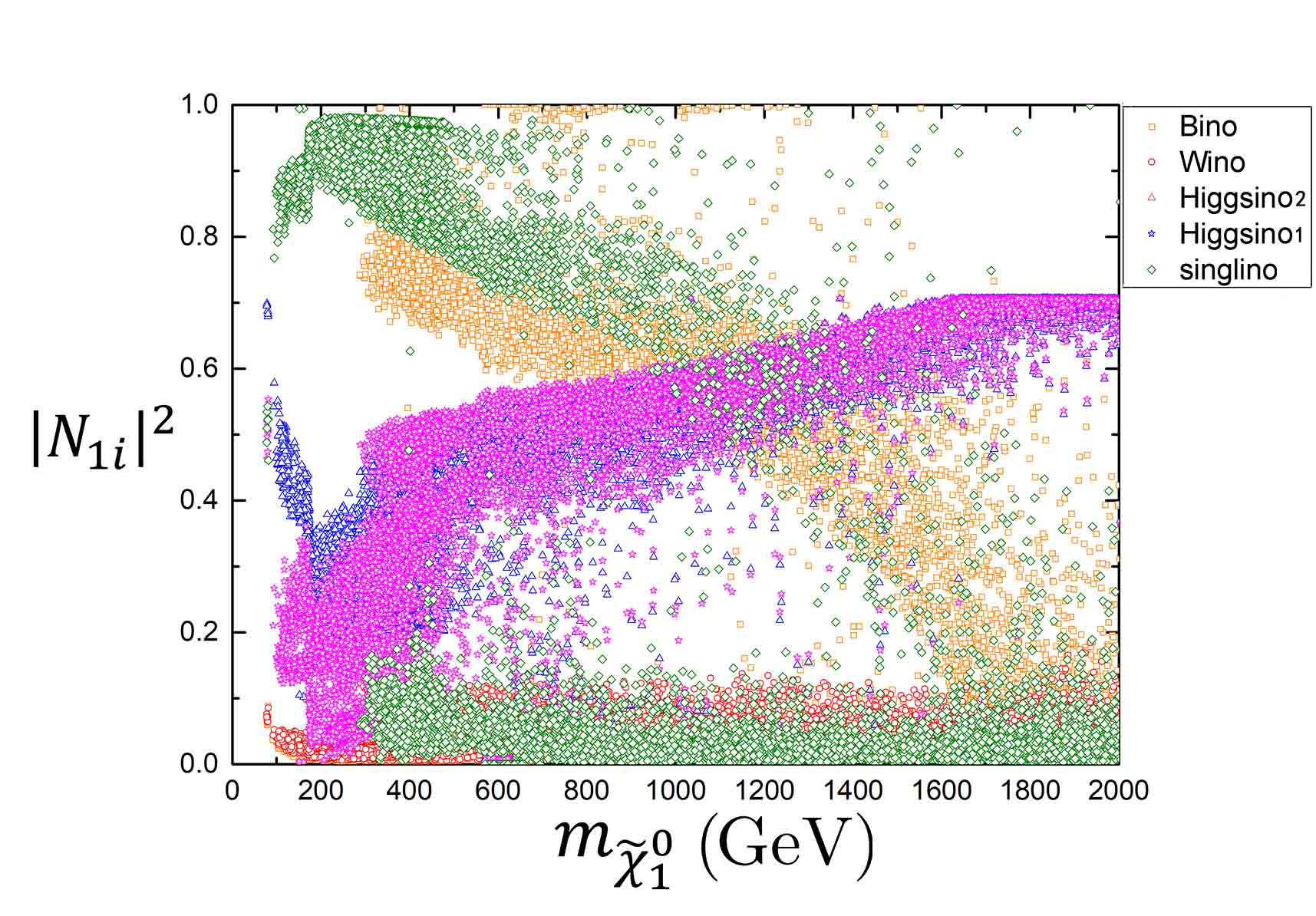}
\includegraphics[width=2.9in]{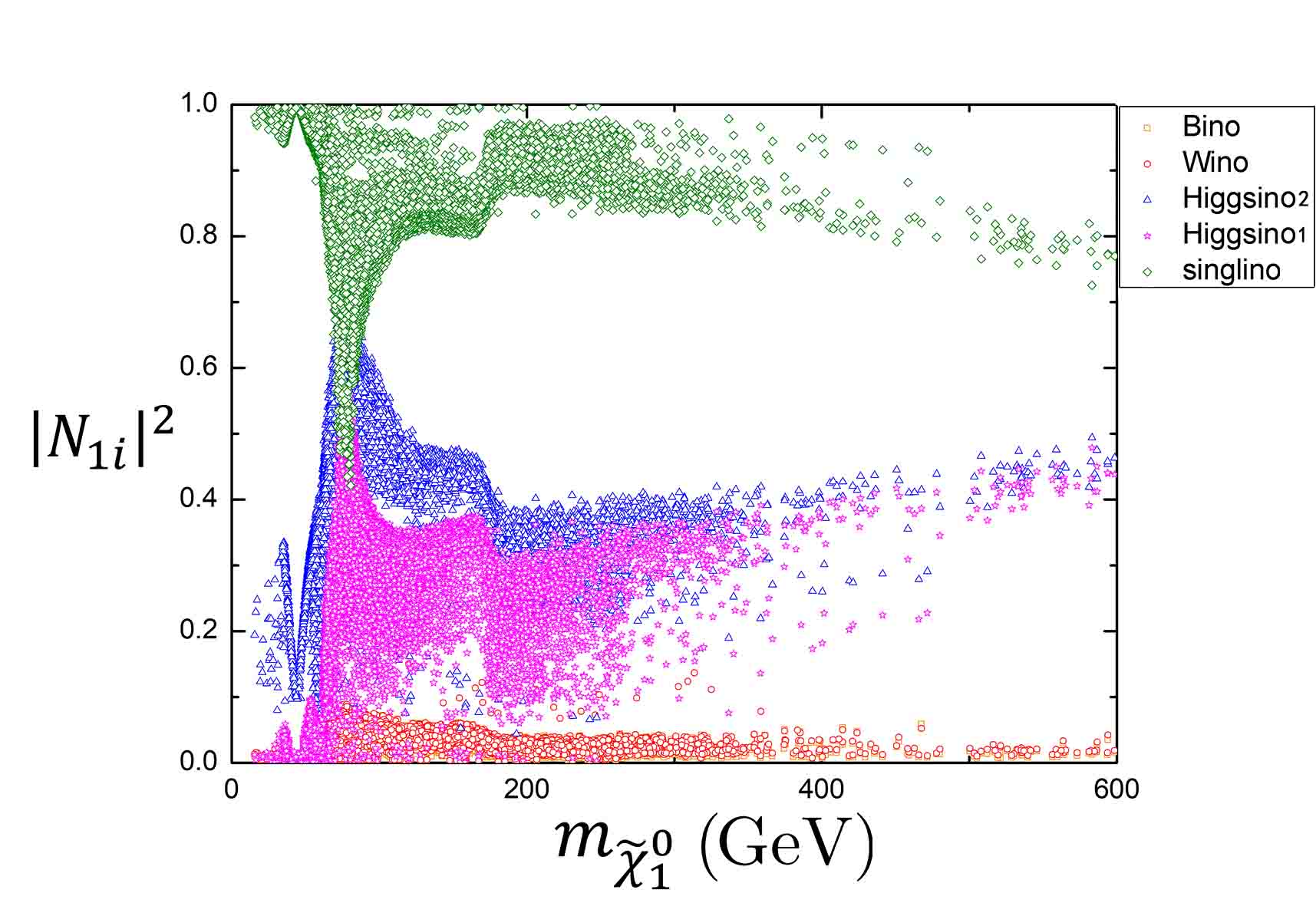}\\
\includegraphics[width=2.9in]{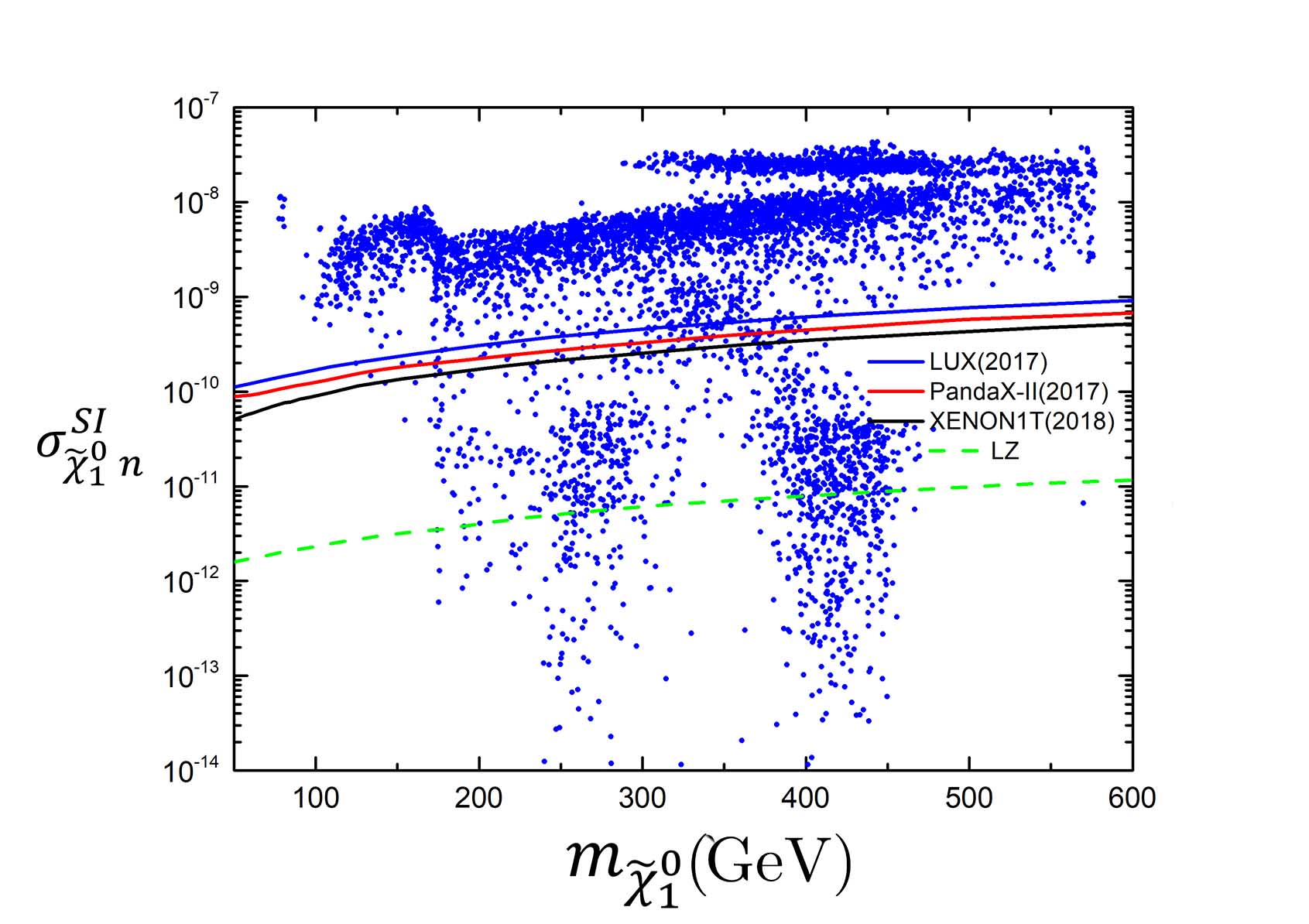}
\includegraphics[width=2.9in]{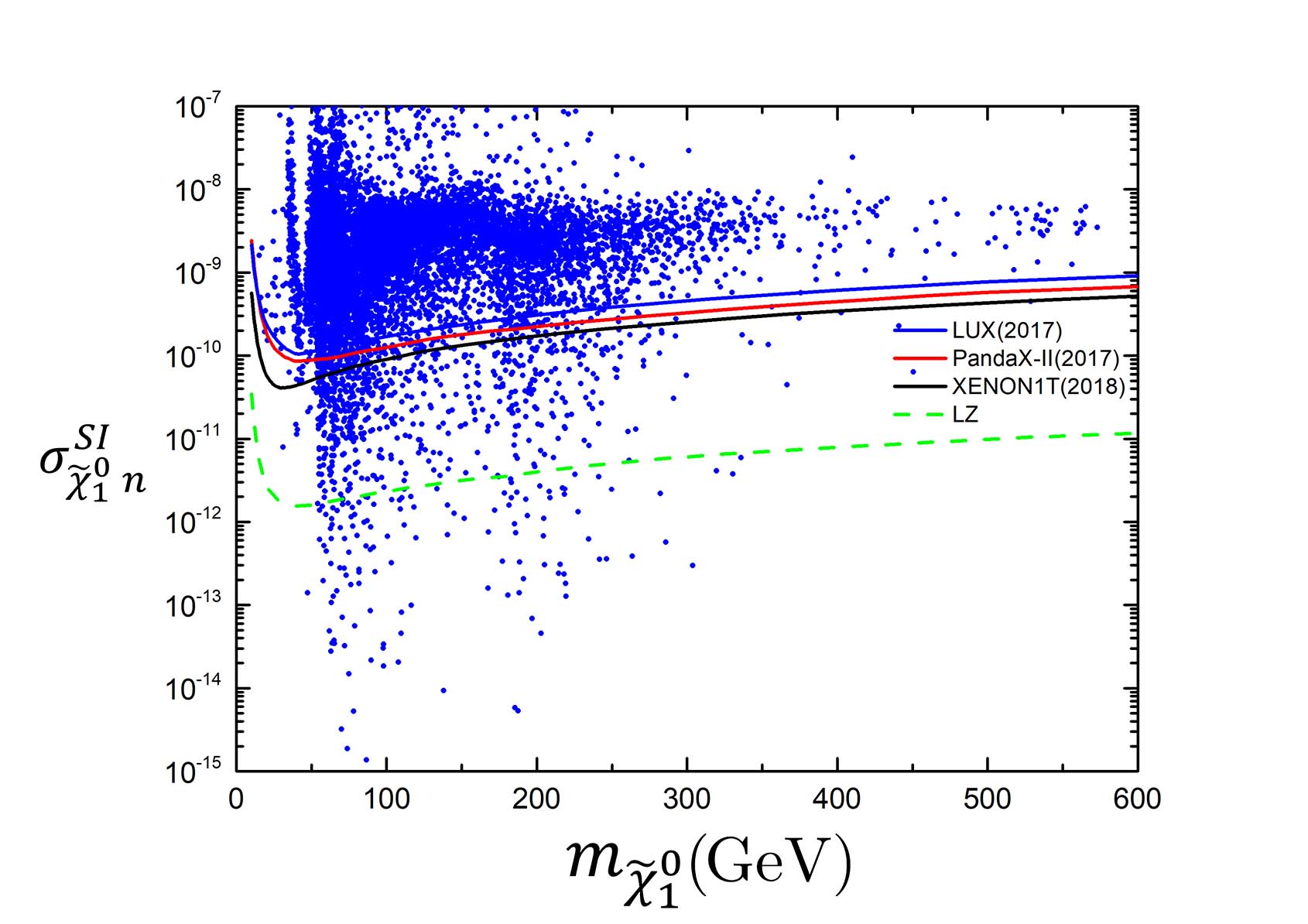}
\end{center}
\vspace{-.5cm}
\caption{ The upper panels show the plots of DM mass vs the DM components for Type A (left panel) or Type B (right panel).
Similarly, the lower panels show the plots of DM mass versus the Spin-Independent(SI)
direct detection bounds. All samples satisfy the constraints from (I-V). }
\label{fig4}
\end{figure}

We know that mixed bino-Higgsino DM is severely constrained by direct detection constraints. For a singlino-dominant DM, the exchange of a light $H_1$ can possibly lead to a large direct detection cross section that will be accessible in the present generation of detectors. It can be seen from the lower panels of Fig.\ref{fig4} that only a small portion of DM parameter space can survive the spin-independent (SI) DM direct detection constraints from the LUX \cite{LUX2016}, PANDAX \cite{PANDAX} and Xenon1T\cite{XENON1T2018}. In fact,  direct detection constrained the DM mass to lie in the range $[120,470]$ GeV for Type {\bf A} and  $[50,400]$ GeV for Type {\bf B}.

\begin{figure}[htb]
\begin{center}
\includegraphics[width=2.9in]{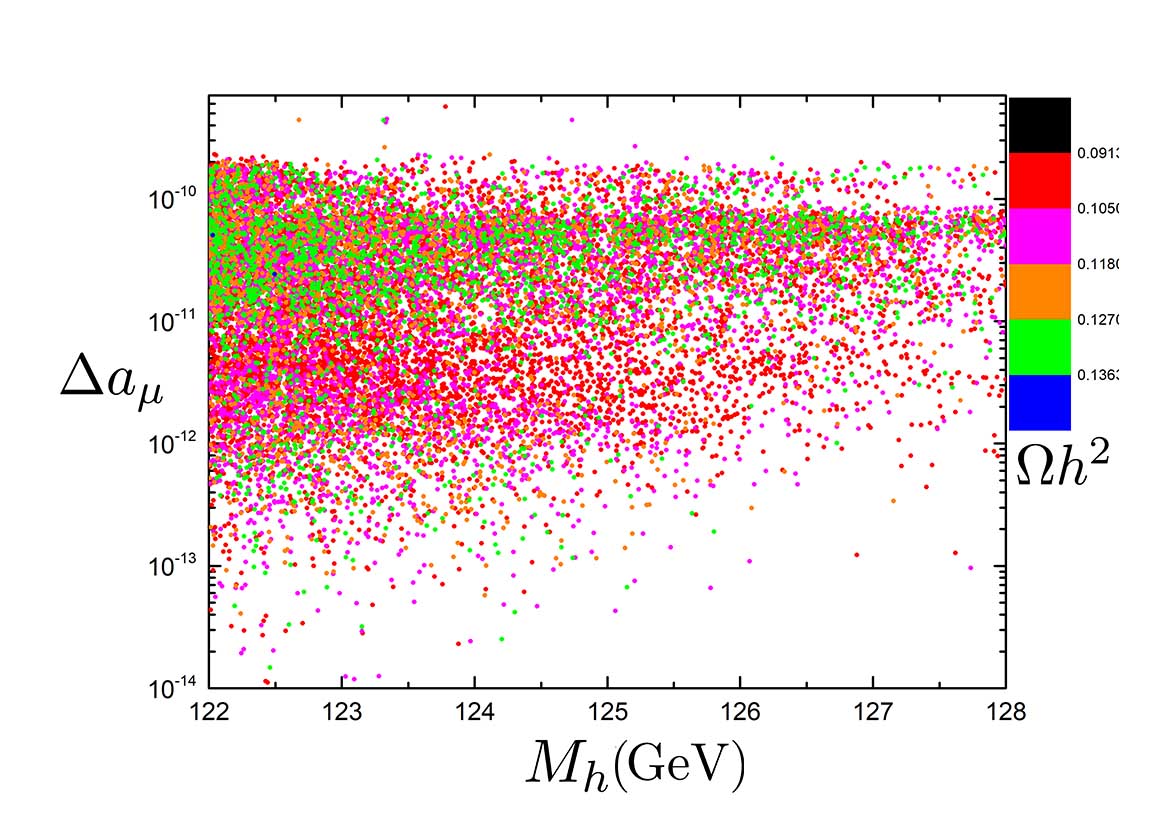}
\includegraphics[width=2.9in]{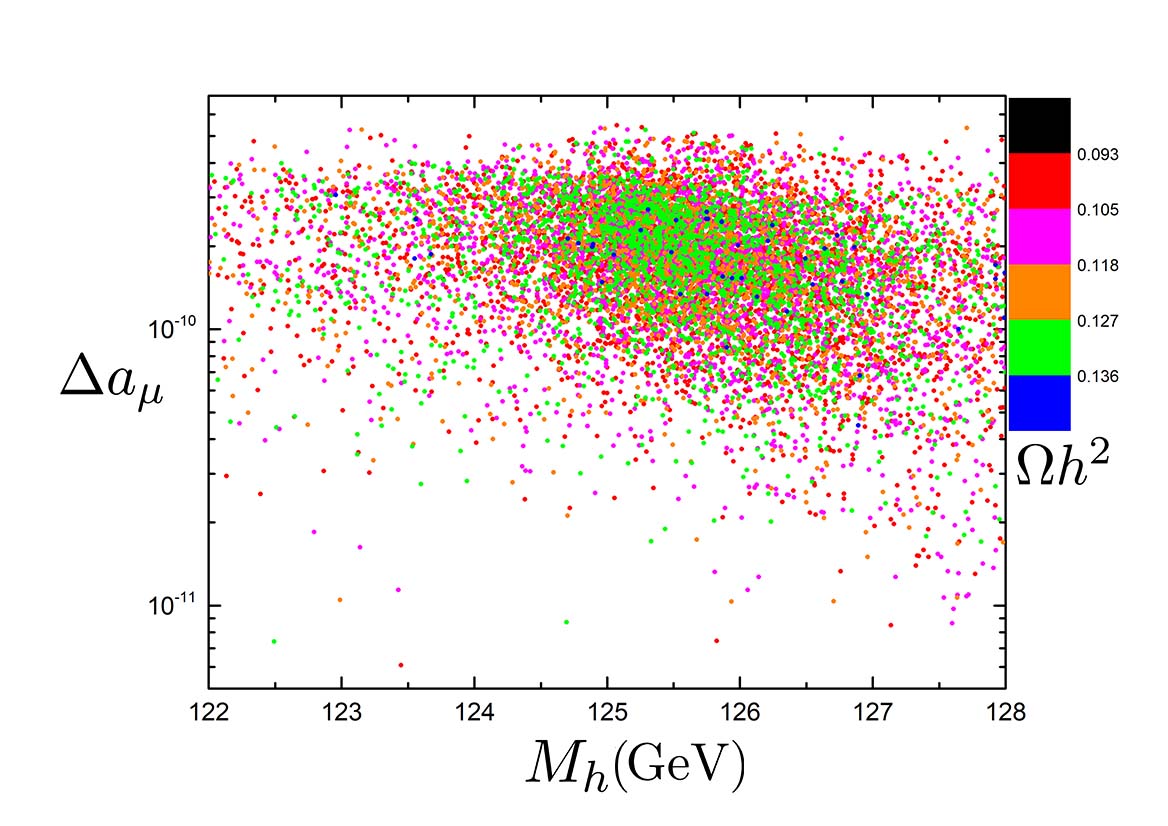}\\
\includegraphics[width=2.9in]{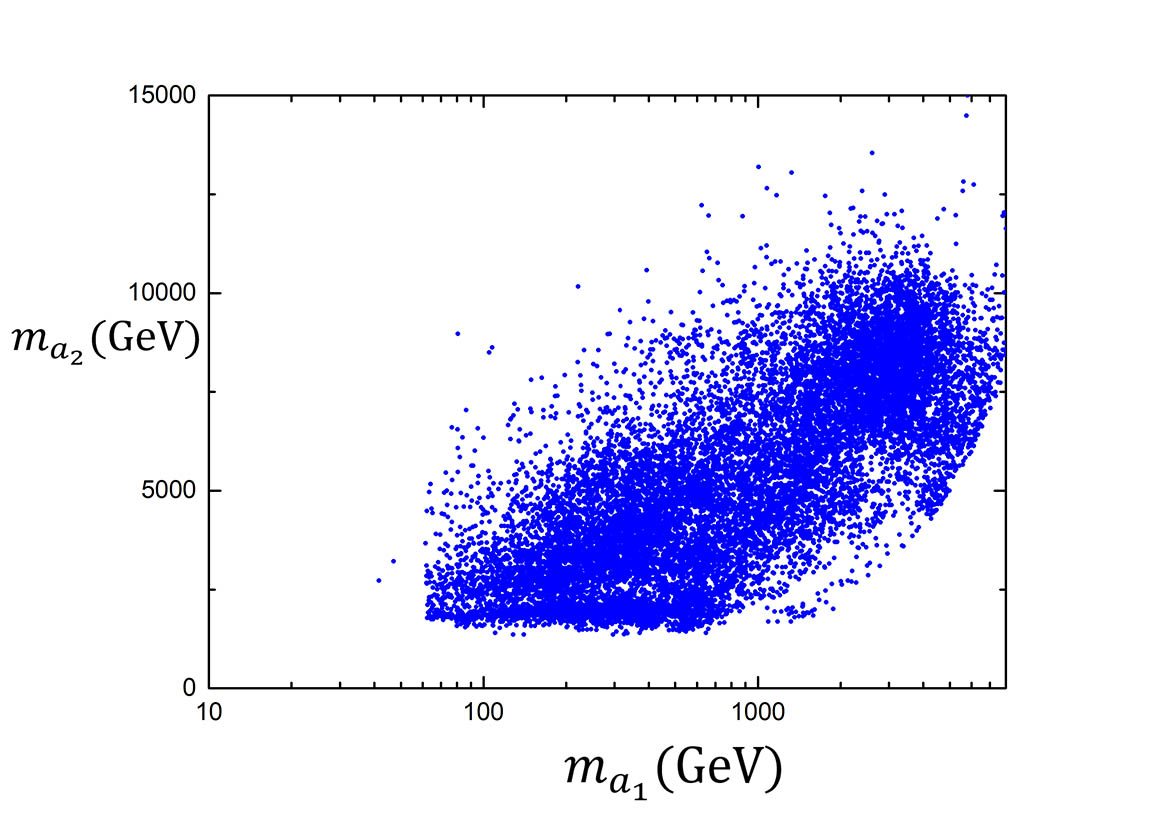}
\includegraphics[width=2.9in]{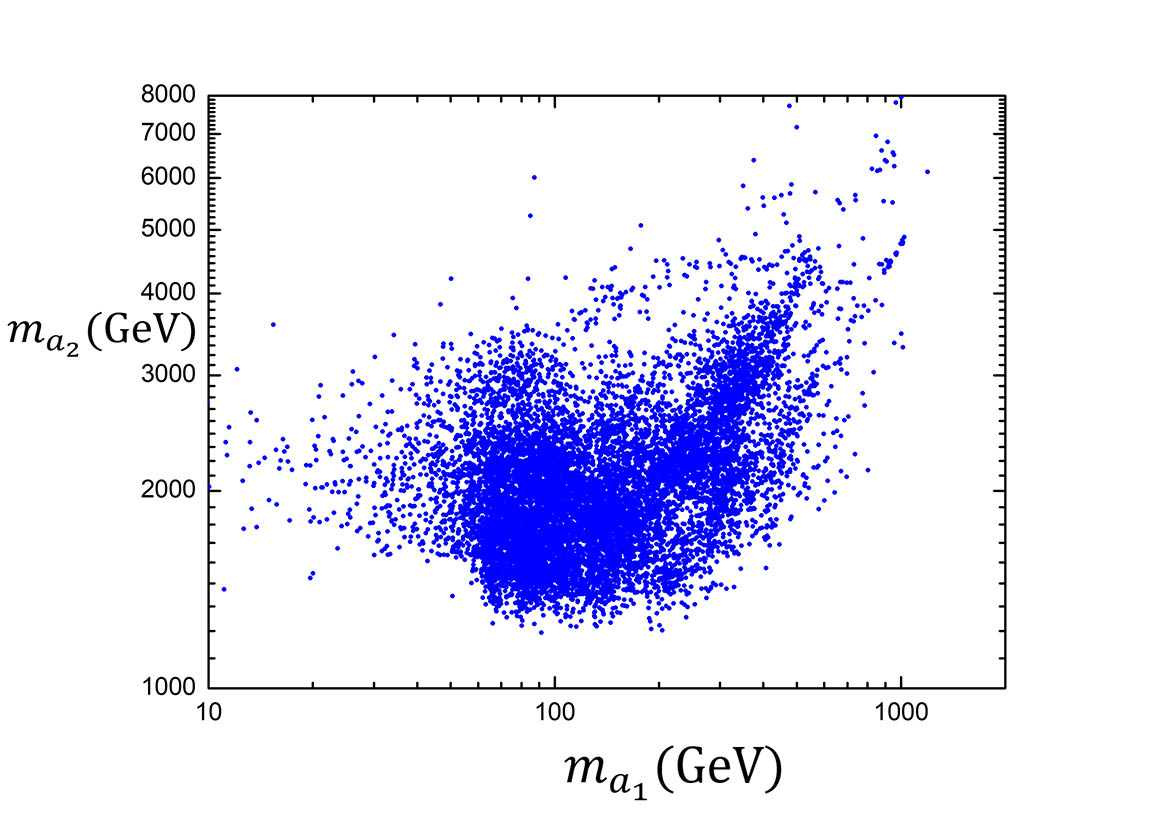}
\end{center}
\vspace{-.5cm}
\caption{ The upper panels show the SUSY contributions to the muon anomalous magnetic moment $\Delta a_\mu$ for Type A(left) and Type B(right).
The lower panels show the corresponding masses of the lightest CP-odd scalar $m_{a_1}$, which can possibly give large two-loop contributions to $\Delta a_\mu$.
All samples satisfy the constraints (I-V). }
\label{fig5}
\end{figure}

    \item  Fig.\ref{fig5} shows the SUSY contributions to the muon $g-2$.
It is known that the required SUSY contributions to $\Delta a_\mu$ can be achieved only if
the relevant sparticles( $\tl{\mu},\tl{\nu}_\mu, \tl{B},\tl{W}, \tl{H}$) are lighter than $600\sim 700$ GeV
for $\tan\beta\sim 10$ in the MSSM \cite{muon:g-2}.
The inclusion of the singlino in the NMSSM can not give sizable contributions to $\Delta a_\mu$
because of the suppressed couplings of singlino to MSSM sector.
Although the two loop contributions involving the Higgs are negligible in the SM, new Higgs bosons in the
NMSSM could have an important impact on $\Delta a_\mu$ if the lightest neutral CP-odd Higgs scalar is very
light \cite{NMSSM:g-2}. In fact, a positive two-loop contribution is numerically more important for a light
CP-odd Higgs being a bit heavier than 3 GeV and the sum of both one-loop and two-loop contributions is
maximal around $m_{a_1}\sim 6$ GeV.
In our scenario, the lightest CP-odd Higgs $a_1$ is bounded to be heavier than 40 GeV and give negligible two-loop contributions
to $\Delta a_\mu$ in Type {\bf A}. The main contribution to $\Delta a_\mu$ is thus similar to that in the MSSM.
In Type {\bf B}, the lightest CP-odd Higgs $a_1$ can lie near 10 GeV and will give important two-loop contributions
to increase $\Delta a_\mu$ to values favored by experiments.
\item   Before we finish our discussions on numerical results, we note that the gauge-modulus
interference contribution for soft scalar masses will play an important role in  phenomenological
studies. In our scenario, such a contribution is non-vanishing only for $S$ which is given in
eqn.(\ref{interference:g-m}). This contribution will possibly change the EWSB condition,
the dark matter relic density and other collider predictions. We show several benchmark points which are affected by the contributions of $\Delta_P^{TX}(m_{\tl{S}^2})$.  Table.\ref{Delta1} shows several benchmark points, which should not survive the various constraints if the gauge-modulus interference contribution is absent. Table.\ref{Delta2}, in contrary, shows several benchmark points which should survive the various constraints if the gauge-modulus interference contribution is absent.

\begin{table}[htbp]
\caption{Benchmark points, which should not survive the various constraints if the gauge-modulus interference contribution is absent. The quantities with mass dimension are in unit of GeV.
Without the gauge-modulus interference contribution, the points will not survive the constraints  because of the constraints shown in $'Reasons'$.}
 \begin{tabular}{|c|c|c|c|}
 \hline
 &Sample I& Sample II& Sample III\\
 \hline
 d&0.326&0.257&0.036\\
 \hline
 $\al$&0.102&0.075&0.115\\
 \hline
 $M_{mess}$&$1.420\tm10^{9}$&$4.823\tm10^{6}$&$4.724\tm10^{7}$\\
 \hline
 $M_0$ &$1278$& $1203$& $1222$ \\
 \hline
 $\la$ &$0.607$&$0.430$&$0.591$\\
 \hline
 $\ka$ &$0.205$&$0.147$&$0.193$\\
 \hline
 $\la_0$&$2.389$&$1.012$&$2.597$\\
 \hline\hline
 $m_{\tl{Q}_{L;1,2}}$&$2041$  &$1452$& $1479$\\
 \hline
 $m_{\tl{Q}_{L,3}}$&$1714$  &$1081$& $1164$\\
 \hline
 $m_{\tl{U}_{L;1,2}}$&$2096$ &$1558$ &$1584$\\
 \hline
 $m_{\tl{U}^c_{L,3}}$&$1640$ &$1096$ &$1236$\\
 \hline
 $m_{\tl{D}^c_{L;12}}$&$2113$&$1565$ &$1598$  \\
 \hline
 $m_{\tl{D}_{L,3}}$&$1908$&$1322$ &$1345$  \\
 \hline
 $m_{\tl{L}_{L;1,2,3}}$&$920.9$&$897.1$ &$891.2$  \\
 \hline
 $m_{\tl{E}_{L;1,2,3}}$&$761.7$&$691.0$ &$661.4$  \\
 \hline
 $A_{\la}$&$1823$&$1523$&$1807$\\
 \hline
 $A_{\ka}$&$-169.2$&$-163.2$&$-95.60$\\
 \hline
 $A_t$&$-2360$&$-2821$ &$-2401$\\
 \hline
 $A_b$&$-3481$&$-3353$ &$-3450$\\
 \hline
 $A_\tau$&$1412$ &$1453$ & $1371$ \\
 \hline
 $M_{\tl{g}}$ &$2971$&$2889$ &$2775$ \\
 \hline
 $\mu_{eff}$ &$646.3$&$200.9$ &$635.2$ \\
 \hline\hline
 $m_{h_1}$&$125.8$   &$125.7$&$126.9 $ \\
 \hline
 $g_\mu-2$&$5.049\tm 10^{-11}$&$2.170\tm10^{-10}$&$6.035\tm10^{-11}$\\
 \hline
 $\Omega_\chi h^2$&$0.114$&$0.118$ &$0.120$\\
 \hline
 $m_{\tilde{\chi}_1^0}$&$442.0$ &$122.6$&$418.5$\\
 \hline
 $\sigma_P^{SI}$&$1.172\tm10^{-11}pb$   &$2.960\tm10^{-12}pb$  &$1.701\tm10^{-12}pb$\\
 \hline\hline
 Reasons & EWSB &EWSB & collider;Higgs mass; $\Omega h^2$ \\
 \hline\hline
 \end{tabular}
\label{Delta1}
\end{table}
\begin{table}[htbp]
\caption{Benchmark points, which should survive the various constraints if the gauge-modulus interference contribution is absent. The quantities with mass dimension are in unit of GeV.
With the gauge-modulus interference contribution, the points will not survive the constraints because of the constraints shown in $'Reasons'$.}
 \begin{tabular}{|c|c|c|c|}
 \hline
 &Sample I& Sample II& Sample III\\
 \hline
 d&1.114&0.307&1.102\\
 \hline
 $\al$&0.056&0.102&2.804\\
 \hline
 $M_{mess}$&$8.188\tm10^{9}$&$4.360\tm10^{12}$&$3.880\tm10^{12}$\\
 \hline
 $M_0$ &$5049$& $2034$& $938.1$ \\
 \hline
 $\la$ &$0.005$&$0.097$&$0.105$\\
 \hline
 $\ka$ &$0.553$&$0.354$&$0.231$\\
 \hline
 $\la_0$&$1.597$&$2.279$&$0.381$\\
 \hline\hline
 $m_{\tl{Q}_{L;1,2}}$&$9087$  &$3853$& $17479$\\
 \hline
 $m_{\tl{Q}_{L,3}}$&$7478$  &$3155$& $15378$\\
 \hline
 $m_{\tl{U}_{L;1,2}}$&$9200$ &$3836$ &$17234$\\
 \hline
 $m_{\tl{U}^c_{L,3}}$&$6789$ &$2756$ &$12746$\\
 \hline
 $m_{\tl{D}^c_{L;12}}$&$9259$&$3867$ &$17223$  \\
 \hline
 $m_{\tl{D}_{L,3}}$&$8366$&$3482$ &$17057$  \\
 \hline
 $m_{\tl{L}_{L;1,2,3}}$&$3648$&$1465$ &$1377$  \\
 \hline
 $m_{\tl{E}_{L;1,2,3}}$&$3249$&$1449$ &$3455$  \\
 \hline
 $A_{\la}$&$4657$&$1511$&$-2737$\\
 \hline
 $A_{\ka}$&$-6511$&$-3614$&$-4573$\\
 \hline
 $A_t$&$-10187$&$-3843$ &$9530$\\
 \hline
 $A_b$&$-11949$&$-4828$ &$6936$\\
 \hline
 $A_\tau$&$6368$ &$2267$ & $-4347$ \\
 \hline
 $M_{\tl{g}}$ &$11819$&$4523$ &$17463$ \\
 \hline
 $\mu_{eff}$ &$439.2$&$288$ &$4998$ \\
 \hline\hline
 $m_{h_1}$&$127.1$   &$123.8$&$123.6 $ \\
 \hline
 $g_\mu-2$&$1.580\tm10^{-11}$&$1.096\tm10^{-10}$&$-1.110\tm10^{-11}$\\
 \hline
 $\Omega_\chi h^2$&$0.022$&$0.010$ &$0.004$\\
 \hline
 $m_{\tilde{\chi}_1^0}$&$458.3$ &$294.5$&$421.0$\\
 \hline
 $\sigma_P^{SI}$&$1.030\tm10^{-10}pb$   &$7.316\tm10^{-10}pb$  &$1.870\tm10^{-12}pb$\\
 \hline\hline
 Reasons & EWSB &EWSB & collider;Higgs mass; $\Omega h^2$ \\
 \hline\hline
 \end{tabular}
\label{Delta2}
\end{table}

\eit

\section{\label{sec-5}Conclusions}
We propose to generate a realistic soft SUSY breaking spectrum for Next-to-Minimal Supersymmetric Standard Model with a generalized deflected mirage mediation scenario,
 in which additional Yukawa and gauge mediation contributions are included to deflect the RGE trajectory. Based on the Wilsonian effective action obtained by integrating out the messengers, the NMSSM soft SUSY breaking spectrum can be given analytically at the messenger scale. We find that additional contributions to $m_S^2$ can possibly ameliorate the stringent constraints from the EWSB  and 125 GeV Higgs mass. Constraints from dark matter and fine-tuning are also discussed.
 The Barbieri-Giudice fine-tuning measure and electroweak fine-tuning measure
  in our scenario can be as low as ${\cal O}(1)$, which possibly indicates that
  our scenario is natural.

\section*{Acknowledgment}
We are very grateful to the referee for efforts to improve our paper. This work was supported by the National Natural Science Foundation of China
(NNSFC) under grant Nos. 11675147, 11775012, 11705093 and 11675242,
by the Young Core Instructor Foundation of Henan Education Department, by Peng-Huan-Wu Theoretical Physics Innovation Center (11747601),
by the CAS Center for Excellence in Particle Physics (CCEPP),
by the CAS Key Research Program of Frontier Sciences,
by a Key R\&D Program of Ministry of Science and Technology of China
under number 2017YFA0402200-04, and by the ARC
Centre of Excellence for Particle Physics at the Tera-scale under grant
CE110001004.

\appendix
\section{The mirage scale in deflected mirage mediation mechanism}
The gaugino mass at scale $\mu$ below the messenger scale can be written as
\beqa
M_i&=& M_0\f{g_i^2(\mu)}{g_i^2(GUT)}+ F_\phi\f{\al_i}{4\pi}\(b_i-d\Delta b_i\)~,
\label{append}
\eeqa
with
\beqa
\f{1}{g_i^2(\mu)}&=&\f{1}{g_i^2(GUT)}+\f{b_i+\Delta b_i}{8\pi^2}\ln\(\f{M_G}{M}\)+\f{b_i}{8\pi^2}\ln\(\f{M}{\mu}\),\nn\\
&=&\f{1}{g_i^2(M_Z)}-\f{b_i}{8\pi^2}\ln\(\f{\mu}{M_Z}\)~.
\eeqa
Here $M_G,M$ denote the gauge coupling unification scale and the messenger scale, respectively.
We will show that apparent $'mirage'$ unification for gaugino masses will still be preserved after the introduction of messenger sector in deflected mirage mediation scenarios. Substituting $\Delta b_i\equiv N$ and the definition ${F_\phi}\equiv (16\pi^2)\al M_0$ into Eqn.(\ref{append}), the gaugino masses can rewrite as
\beqa
M_i&=& M_0\f{g_i^2(\mu)}{g_i^2(GUT)}+ \al M_0 \(b_i-d N\) g_i^2(\mu)~,\nn\\
&=& \[{M_0}-\al M_0 d N {g_i^2(GUT)}\]\[1-\f{b_i+N}{8\pi^2}g_i^2(\mu)\ln\(\f{M_G}{M}\)-\f{b_i}{8\pi^2}g_i^2(\mu)\ln\(\f{M}{\mu}\)\]\nn\\
&+&\al M_0 b_i g_i^2(\mu)~,\nn\\
&\approx&\({M_0}-\al M_0 d N {g_i^2(GUT)}\)\[\left(1-\f{N}{8\pi^2}g_i^2(M_{G})\ln\(\f{M_G}{M}\)\right)-\f{b_i}{8\pi^2}g_i^2(\mu)\ln\(\f{M_G}{\mu}\)\]\nn\\
&+&\al M_0 b_i g_i^2(\mu)~,\nn\\
&\equiv& K_0\[c_0-\f{b_i}{8\pi^2}g_i^2(\mu)\ln\(\f{M_G}{\mu}\)\]
+\al M_0 b_i g_i^2(\mu)~,
\eeqa
with
\beqa
K_0&\equiv& M_0-\al M_0 d N {g_i^2(GUT)}~,\nn\\
c_0&=& 1-\f{N}{8\pi^2}g_i^2(M_{G})\ln\(\f{M_G}{M}\)~.
\eeqa
So we can see that mirage unification for gaugino masses will be satisfied at the scale $\mu$ determined by
\beqa
\ln\(\f{M_G}{\mu}\)=\f{8\pi^2 \al M_0}{K_0}~.
\eeqa
with the mirage unification values for gaugino masses as
\beqa
M_i(\mu_{mirage})=K_0c_0~.
\eeqa

\end{document}